\documentclass[preprint,12pt]{elsarticle}

\usepackage{amssymb}
 \usepackage{amsthm}

\usepackage{lineno}

\usepackage[caption=false,font=footnotesize]{subfig}

\usepackage{algorithmicx}
\usepackage[]{algpseudocode}
\usepackage{array}
\usepackage{amsmath, amsfonts}
\usepackage{listings}
\usepackage{xcolor}
\usepackage{url}

\usepackage{tikzexternal}
\tikzsetfigurename{jsa-2020-figure}

\newcolumntype{d}[1]{>{\centering\arraybackslash}m{#1\columnwidth}}

\newtheorem{problem}{Problem}
\newtheorem{theorem}{Theorem}
\newtheorem{definition}{Definition}

\newtheorem{lemma}{Lemma}

\lstdefinelanguage{strl}{
  belowcaptionskip=1\baselineskip,
  breaklines=true,
  xleftmargin=2.1em,
  showstringspaces=false,
  basicstyle=\footnotesize\ttfamily,
  keywordstyle=\bfseries\color{brown},
  commentstyle=\itshape\color{black!70!white},
  morekeywords={module,var,await,abort,present,then,else,nothing,if,loop,end,emit,pause,and,
    		   integer,when,sustain,case,do,float},
  numbers=left,
  tabsize=1,
  morecomment=[l]{\%},
  escapeinside={@}{@},
}

\journal{Journal of Systems Architecture}

\begin{document}

\begin{frontmatter}



\title{Online Cycle Detection for Models with Mode-Dependent Input and Output Dependencies}


\author[a]{Heejong Park}
\author[a]{Arvind Easwaran}
\author[b]{Etienne Borde}

\address[a]{Nanyang Technological University, Singapore}
\address[b]{LTCI, T\'{e}l\'{e}com Paris, Institut polytechnique de Paris, France}

\begin{abstract}

In the fields of co-simulation and component-based modelling, designers import models as building blocks to create a composite model that provides more complex functionalities. 
Modelling tools perform instantaneous cycle detection (ICD) on the composite models having feedback loops to reject the models if the loops are mathematically unsound and to improve simulation performance. 
In this case, the analysis relies heavily on the availability of dependency information from the imported models. 
However, the cycle detection problem becomes harder when the model's input to output dependencies are mode-dependent, i.e. changes for certain events generated internally or externally as inputs.
The number of possible modes created by composing such models increases significantly and unknown factors such as environmental inputs make the offline (statical) ICD a difficult task. 
In this paper, an online ICD method is introduced to address this issue for the models used in cyber-physical systems. The method utilises an oracle as a central source of information that can answer whether the individual models can make mode transition without creating instantaneous cycles. 
The oracle utilises three types of data-structures created offline that are adaptively chosen during online (runtime) depending on the frequency as well as the number of models that make mode transitions. 
During the analysis, the models used online are stalled from running, resulting in the discrepancy with the physical system. The objective is to detect an absence of the instantaneous cycle while minimising the stall time of the model simulation that is induced from the analysis. The benchmark results show that our method is an adequate alternative to the offline analysis methods and significantly reduces the analysis time.
\end{abstract}




\end{frontmatter}


\section{Introduction}
Today, there exist a plethora of modelling tools for capturing the behaviour of physical systems. Many of these tools provide a set of well-established libraries that can be reused for constructing more complex systems. 
Reusability is a significant part of many engineering tasks where designers can rely on the correctness of individual components, which have been used and tested extensively by others. A library of well-defined components, however, does not always guarantee the correctness of a model made of such components. For example, one of the useful techniques in modelling is a feedback loop where output events of a system are routed back as its inputs and create cyclic dependencies. Nevertheless, creating feedback loops in an undisciplined manner can result in \textit{instantaneous cycles}. It refers to the situation when the cause of input and output events for a model are interdependent with each other that can make composite models (models that are made of other smaller models) mathematically unsound. This leads to a simulation that diverges from the operations of the physical system and increases the simulation time of the composite model~\cite{broman2013determinate}.

The classical approach for checking the absence of instantaneous cycles in a composite model is to reject any feedback loops that are not broken by a unit delay. A delay is typically introduced in a model via components such as an integrator and unit delay components whose output events do not depend on their present or future input events at any time instance. Thus, a modelling tool performs instantaneous cycle detection (ICD) during design time (offline) based on the model composition and \textit{statically} rejects models with feedback loops whose computational dependency cannot be solved mathematically. Yet, in the environment where models are imported as third-party components~\cite{gomes2018co}, detecting cycles in a composite model can be difficult. This is mainly because the library vendors hide the internal implementation of the models due to Intellectual Property (IP) issues, prohibiting accurate analysis of the design. Another approach consists in rejecting any model compositions that structurally appear as feedback loops on a top-level or automatically inserting a delay component in every loop. Yet, such methods can potentially reject models that are acyclic when the models are flattened to the lowest level of the hierarchy~\cite{lublinerman2009modular}. 

ICD becomes more difficult when the input to output dependencies of models dynamically change based on their \textit{internal mode} (like in hybrid automata). There is a body of work~\cite{matsikoudis2015fixed,zhou2008causality,broman2013determinate} which employ \textit{causal dependency} information of individual sub-model's input and output ports to check instantaneous cycles in a composite model. Most often this information is provided by the library vendors or the tools that generate models so that the tool that performs the assembly of components perform ICD. For a composite model consisting of mode-dependent sub-models, the analysis should verify that all reachable modes exclude instantaneous cycles. However, statically filtering out such erroneous modes in a composite model through a reachability analysis is typically an undecidable problem~\cite{zhou2008causality} for white-box models and even infeasible for component-based models. 
These components are typically grey-box models whose mathematical behaviours are hidden and only partial information, such as input to output dependency, is exposed to their external environment.

In this paper, we instead tackle the mode-dependent ICD problem via an online method that provides an efficient way to reduce the \textit{stall time} of a component-based model simulation due to the analysis. 
The stall time is the duration that the models are blocked from execution due to housekeeping works such as ICD.
We target areas of cyber-physical systems including but not limited to: model driven control systems where the models continuously synchronise with the physical system such that the results of the model simulation influence operations of the physical system or vice-versa. 
Therefore, a large stall time causes an unwanted deviation of models from their represented physical counterparts.
To our knowledge, there is no literature to this date that tries to address the overhead due to the \textbf{online} ICD for \textbf{mode-dependent} input to output dependencies in grey-box component-based models. In particular, our method employs an \textit{oracle}, which can provide `yes' or `no' answers to individual models requesting a mode change authorisation. Therefore, our approach does not need to verify all reachable modes combinations from the model composition but detects an instantaneous cycle online whenever the models are about to change their modes. The key problem in our approach is to reduce the stall time required for the oracle to decide if the requested mode changes are acceptable therefore does not create discrepancies between the models and the physical system.
The detailed contributions of this paper are:
\begin{enumerate}
\item An online ICD method for Cyber-Physical Systems (CPS) consisting of component-based models whose input and output dependencies are mode-dependent.

\item An adaptive method to reduce the \textit{stall time} (time consumed by ICD algorithm) of models that synchronise with physical systems in a real-time manner and thus can be used in various CPS applications.

\item A set of benchmark results and a case study that shows the applicability of our method in the fields of industrial manufacturing systems.
\end{enumerate}

The rest of this paper is organised as follows: Section~\ref{sec:prob} defines the problem that we address in this paper. Section~\ref{sec:rel} enumerates the related research on ICD followed by the introduction of the workpiece sorting system as our motivating example in Section~\ref{sec:mot}. Preliminary background is presented in Section~\ref{sec:bg}. The methodology of our online ICD is introduced in Section~\ref{sec:met}. \mbox{Sections~\ref{sec:exp}} and \ref{sec:wp-case} present a set of experimental results that evaluate the performance of the analysis in various settings. A set of use-case scenarios of our technique is presented in Section~\ref{sec:dis} through existing real-world examples. Finally, conclusion and future work are given in Section~\ref{sec:concl}.

\section{Problem Definition}\label{sec:prob}

\begin{figure}[t!]
\centering
\includegraphics[width=0.71\linewidth,page=1]{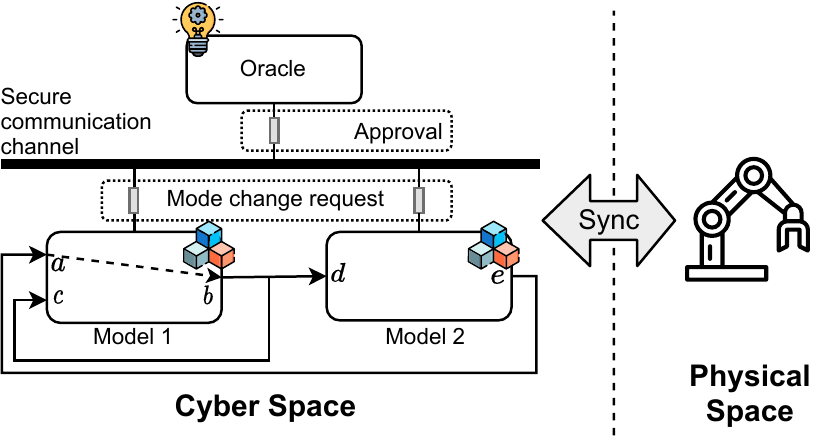}
\caption{Oracle-based instantaneous cycle detection}
\label{fig:overview}
\end{figure}

A graphical overview of the proposed oracle-based online ICD is shown in Fig.~\ref{fig:overview}. We consider models as a set of components with their input and output ports interconnected with each other to exchange data during the simulation. Furthermore, the models simulated in the cyber space synchronise operations with their physical counterparts in the physical space.
In the cyber space, the arrows indicate a data dependency from output to input \textit{between} the models 1 and 2 or vice-versa \textit{within} the models. For example, to generate output data from the output port $ b $ from the model 1, it requires an input value from port $ a $, which is generated from the output port $ e $ from the model 2. When there is no arrow incident to an output port, for example $ e $, it can generate data without the need for the input port values to be resolved. The composition of these models shown in the figure is acyclic because there are no paths along the input and output dependencies that create a cycle, i.e. creating an instantaneous cycle. In addition, we also consider the models that can change the input to output dependencies along with their operating modes. For example, creating a dependency between the ports $ d $ and $ e $ upon a mode change in model 2. In this case, an instantaneous cycle is created for the path $ e\to a\to b\to d \to e$. Our objective is to quickly detect the creation of such a cycle during runtime due to mode changes in the models. To achieve this, we assume the models make \textit{mode change requests} (MCR) to the central oracle, which monitors the creation of the instantaneous cycles before the mode change can happen. Each MCR from a model only contains local dependency information and the oracle ensures this IP-sensitive information is not shared among the models. The oracle checks whether the mode change is valid and responds with an approval message back to the models within a finite time bound. We call this bound \textit{stall time} ($ t_{stall} $) because the models must stall and not proceed to the next simulation step until they receive a response from the oracle for their MCR requests. Stalling models do not require additional features in these models since their execution is controlled by an external entity, e.g. main loop, that coordinates progression of the model's time. In this work, we consider grey-box, component-based models that expose their current mode's input-to-output dependency information to the environment. Here we define the problem as follows.

\begin{problem}
Develop an online ICD technique that minimises the stall time for grey-box, component-based models whose input and output dependencies can change during runtime depending on their internal modes.
\end{problem}

When the oracle detects an instantaneous cycle from the MCR requests, it rejects the request from the model that creates the loop. Such rejection should not result in a fault in an on-going cyber-physical operation. Therefore, the corresponding model makes a transition to a \textit{safe mode}, which is specified by a designer beforehand and guaranteed not to create an instantaneous cycle. These safe modes would require additional logic in models; however, we think this is a little addition to the models that support multi-mode features which we target in this paper. We also assume the models expose a limited set of safe modes and their combinations with other models are checked statically offline. The size of such combination is typically much smaller than the entire combination of all possible modes. Models that made a transition to a safe mode can return to a normal operating mode in the same way they make an MCR to the oracle. In this way, the analysis can be performed independently by a trusted entity (oracle) and models are only required to provide changes in the local input-to-output dependency information for each mode change.

\section{Related Work} \label{sec:rel}

The problem of detecting cyclic dependencies is a subset of \textit{causality analysis} problem, which can be found in formal modelling literature. Authors in \cite{zhou2008causality} introduced \textit{dependency algebra} to formulate the causality problem for a network of actors whose input and output dependencies are fixed. Dependency algebra is implemented in Ptolemy~II~\cite{lee2014constructive} to check the instantaneous cycle via solving the algebraic equation, which is derived from the composition of actors. Their method checks if every \textit{simple cycle} in the network of actors is broken by the delay component (e.g. the integrator block). Finding all simple cycles, however, requires traversals of the whole actor network as many times as there are cycles in the network. Ptolemy~II also provides an option to support the mode-dependent online ICD where only input-to-output dependencies in active modes are considered for the analysis. However, this option still employs the method introduced in~\cite{zhou2008causality} and thus has the same runtime complexity as solving algebraic equations for the actors with fixed modes. On the other hand, we are introducing an adaptive technique to efficiently detect the mode-dependent instantaneous cycles that is faster than finding simple cycles on the whole network.

Authors in \cite{hofbaur2006causal} introduced a causal analysis method for concurrent hybrid automata. Similar to our approach, their approach tries to detect causality cycles online. They introduced a notion of \textit{compatibility} where two automata are compatible if no output variables are shared in any mode combinations. Nevertheless, checking such compatibility is an expensive task when the number of all possible mode combinations grows exponentially. Our method, on the other hand, does not require checking of all possible mode combinations amongst concurrently running automata, therefore more amenable in the online setting.
Work on the structural analysis of multi-mode differential algebraic equation (DAE) systems is presented in~\cite{benveniste2017structural}. The authors showed an example of modelling a clutch in a car between engaged and released modes. The approach is based on the analysis on the equation (program) level which is hard to directly apply to our case that utilises input and output dependencies on the component level.

The problem of cycle detection especially in a dynamically changing graph dataset has been applied in a variety of applications. In~\cite{qiu2018real}, authors introduced a technique to find cycles in a large-scale graph that satisfy both length and some attribute constraints. The technique is deployed at Alibaba in an e-commerce system to monitor fraudulent activities upon cycle detection. Online cycle detection technique is also used in the pointer analysis in a program \cite{pearce2004online} as well as in a distributed deadlock detection algorithm \cite{lee1992distributed}. In this paper, on the other hand, we introduce the dynamic cycle detection problem in the domain of component-based modelling and present an efficient technique by classifying the types of mode changes within the models.

We foresee our approach can be applied in a variety of applications such as checking dependencies in a task mapping scenarios~\cite{huang2020hda}, dependency reduction algorithms~\cite{kim2019data} and fault detection algorithms that have task execution dependencies as constraints~\cite{jiang2020design}.

\section{Motivating Example} \label{sec:mot}

\begin{figure}[t!]
\centering
\includegraphics[page=2,width=0.75\linewidth]{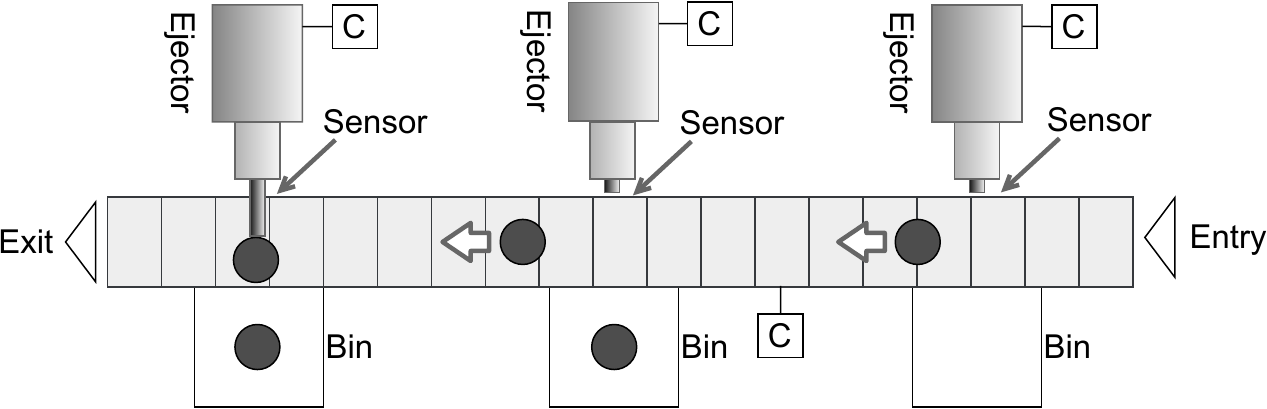}
\caption{A workpiece sorting system}
\label{fig:ex}
\end{figure}

\begin{figure}[t!]
\centering
\includegraphics[page=3,width=0.65\linewidth]{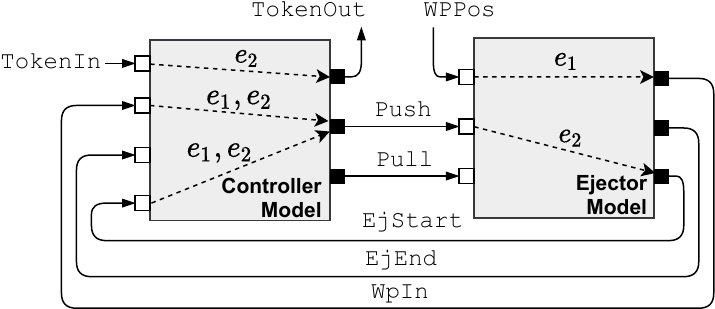}
\caption{An ejector system model consisting of a controller and a plant that runs in the cyber space}
\label{fig:ex2}
\end{figure}

Consider an example of an industrial automation system shown in Fig.~\ref{fig:ex}. It consists of a set of ejectors that extend to place workpieces into one of the bins in front of them. The linear conveyor belt can transfer workpieces to the next workstation via an exit point if they are yet to be considered as final products.
Each machine in the system is controlled by its controller (annotated with `C') which makes them a closed-loop system. In addition, we assume the case where both the controller and the machine (plant) models run concurrently in real-time in the cyber-space. Without further going into implementation details of each model, Fig.~\ref{fig:ex2} illustrates the top-level components of the controller and the ejector, which are interconnected via input (white rectangle) and output (black rectangle) ports. The semantics of communication between these two models are depending on their model of computation (MoC). In this paper, we assume these models are based on the synchronous reactive (SR) MoC~\cite{lee2007leveraging} with an extension of continuous time, where discrete events between these models are \textit{instantaneously} propagated via input and output ports within the same simulation time step called a \textit{tick}. Fig.~\ref{fig:strl-wp} shows the implementation of the controller and plant of the workpiece sorting system in Fig.~\ref{fig:ex2} in an SR language called Esterel~\cite{berry1992esterel}.

\begin{figure}[t!]
\begin{lstlisting}[language=strl]
[ % Controller Model @ \label{c-1} @
	loop 
		present TokenIn then
			present notFull then % Mode-1 (Push) @ \label{c-9} @
				await [EjStart and WpIn];  @ \label{c-4} @
				abort sustain Push when EjEnd;  @ \label{c-5} @
				await EjEnd; 
				abort sustain Pull when EjStart; 
			end present; @ \label{c-9.1} @
			emit TokenOut;
		end present;
		pause
	end loop
] || [ % Plant Model @ \label{c-2} @
	var pos : float in
		loop
				present notFull then % Mode-1
					present @ \label{c-6} @
						case Push do pos := integrate(pos, 1)
						case Pull do pos := integrate(pos, 0)
					end present;
					% 1. Check the location of WP (WpPos) @ \label{c-7} @
					% 2. Emit EjEnd, EjStart based on pos @ \label{c-8} @
				else % Mode-2 @ \label{c-12} @
					present Push then @ \label{c-10} @
						emit EjStart % Cycle
					end present @ \label{c-11} @
				end present; @ \label{c-13} @
				pause
		end loop
	end var
] @ \label{c-3} @
\end{lstlisting}
\caption{Esterel implementation of the controller and plant models for the workpiece sorting system.}
\label{fig:strl-wp}
\end{figure}

The controller model has two outputs \texttt{Push} and \texttt{Pull} that push and pull the ejector to move the workpiece into the bin. Feedback signals \texttt{EjStart} and \texttt{EjEnd} provided by the ejector model indicate if the ejector has been completely retracted or extended, respectively. The input \texttt{WpPos} is a continuous variable that indicates the locations of the workpieces derived from the conveyor belt model (not shown in the figure). \texttt{WpIn} is a sensory input to the controller that triggers it to issue a push signal to the ejector. \texttt{TokenIn} and \texttt{TokenOut} are connected between adjacent controllers to implement a ring-token, which is used to evenly distribute workpieces into the bin. In Fig.~\ref{fig:strl-wp}, the controller (lines \ref{c-1}-\ref{c-2}) and ejector (lines \ref{c-2}-\ref{c-3}) models are executed concurrently with each other indicated by the synchronous parallel operator \texttt{||} (line~\ref{c-2}). In this program, we implement modes using the conditional \texttt{present} statement where each branch indicates a single mode.

These models have several distinct modes to adapt to different operating scenarios. The cyclic dependency problem occurs due to the feedback loops such as \texttt{EjStart}, \texttt{EjEnd} and \texttt{WpIn} as shown in this example. In many cases, it is not clear if these loops are instantaneous cycles in this top-level composition. Indeed, it depends on the internal implementation of each model. Fig.~\ref{fig:ex2} shows an example of input to output dependencies for these models where $ e_{i} $ denotes a dependency in a mode $ i $. When the controller is in mode 1, the output \texttt{Push} depends on both \texttt{EjStart} and \texttt{WpIn} because the controller can only trigger the operation when the ejector is fully retracted \textit{and} the workpiece is detected via a sensor. This dependency is shown in Fig.~\ref{fig:strl-wp} at line~\ref{c-4} using the Esterel statements \texttt{await} that blocks the controller until both \texttt{EjStart} and \texttt{WpIn} become `present' before outputting the signal \texttt{Push} at line \ref{c-5}.
The ejector model does not require \texttt{Push} to generate \texttt{EjStart} since the dynamics of the eject operation breaks this dependency, i.e. an integrator block in the model breaks the dependency. This is shown at lines \ref{c-6}-\ref{c-8} in Fig.~\ref{fig:strl-wp} (some parts are omitted for brevity) that is performed solely on the value of \texttt{pos}. Therefore the model is acyclic in this mode\footnote{Note that dependencies between \texttt{Pull} and \texttt{EjEnd} are also similar but omitted in the figure for the sake of conciseness}. 

We can introduce an alternative mode for these models when the bins are full so that a workpiece cannot be pushed by the ejectors. In this scenario, when the controller tries to push the workpiece via \texttt{Push}, \texttt{EjEnd} never becomes true since the bin is fully occupied. Instead, \texttt{EjStart} becomes false immediately until the controller stops issuing \texttt{Push}. In this case, we created an instantaneous cycle since \texttt{EjStart} is now immediately dependent on \texttt{Push}, i.e. the path where edges of mode 2, noted $ e_2 $, are created in Fig.~\ref{fig:ex2}. In Fig.~\ref{fig:strl-wp}, this dependency is shown at lines~\ref{c-10}-\ref{c-11} where \texttt{EjStart} is only emitted if the signal \texttt{Push} is present. The output of the Esterel compiler (v5.92) upon compilation of this code
indicates there exist a cycle between these signals and the compiler cannot generate statically scheduled code. If we closely inspect the code in Fig~\ref{fig:strl-wp}, such cycle cannot be created if the guarded signal \texttt{notFull} stays present while the control-flow of the controller model is within lines~\ref{c-4}-\ref{c-5}, i.e. the plant model does not enter the lines~\ref{c-10}-\ref{c-11} that creates a cycle. However, such case could not be guaranteed at compile time and the program is rejected for code generation. Therefore, static analysis can still reject models that are not cyclic based on the active modes. Furthermore, it would be even harder to detect cycles created due to the composition of models with multiple modes, for example when the controller and plant models are separately developed in component-based modelling environment as shown in Fig.~\ref{fig:ex2}. This example motivates the need for online technique that dynamically checks the absence of instantaneous cycles.

The controllers pass a token in a circular fashion to select which bin to store an incoming workpiece. When a bin is full, the corresponding controller switches mode in which the dependency $ e_{2} $ between \texttt{TokenIn} and \texttt{TokenOut} is created to immediately pass a token to the next controller. Therefore an instantaneous cycle can also be created in this case such that all controllers pass the token to each other indefinitely. The frequency of the mode changes for the ejector models can cause a lag in the model simulation time due to additional analysis times and result in discrepancies between the models and its physical counterparts.

Offline analysis of the instantaneous cycle becomes more difficult when the system is larger and the total number of modes among the models is significant, especially when the models exhibit concurrent behaviours. Moreover, the analysis has to introduce many pessimistic assumptions when the internal implementations of the model's behaviour are hidden due to IP restrictions. One possible way to mitigate this problem is to perform the analysis during runtime on every mode change and provide safe mode transitions, which are known to be correct, upon a ICD. The safe mode transition is required in the applications where the models are synchronised with the physical system as shown in Fig.~\ref{fig:overview}, to prevent unwanted events, such as physical damages in the system, that cannot be undone. One major advantage of this approach is that there is no need to filter any erroneous modes that require many assumptions which may not be true. Nevertheless, this runtime analysis introduces an unwanted \textit{stall time} that blocks the whole composite model from execution, thus creating a discrepancy such as a delay in the operation between the model and the physical system. In this work, we are introducing a method to reduce such stall time.

\section{Background}\label{sec:bg}

\subsection{Composite Model Structure and Synchronous Reactive Model of Computation}\label{ssec:sr}
A composite model $ C $ in our modelling approach is a tuple $ \langle M,I,O,\allowbreak S,T \rangle $ where $ M $ is a set of component-based models, $ I $ is a set of input ports, $ O $ is a set of output ports and $ S : O\times I $ is a set of interface signals that describe dependencies from output to input ports. A signal is a status and a value pair $(s_{st},s_v) $ where $ s_{st}\in\{0,1,\bot\} $, $ s_v\in\mathbb{R} $. A sequence of \textit{ticks} $ T = \{(n,r) \mid n\in\mathbb{N}, r\in\mathbb{R},n \geq 0\wedge r\geq0\} $ is a shared time among all models within the same composite model. We extend the logical time $ n $ in the traditional SR MoC with a real number $ r $ to describe continuous time models such as Ordinary Differential Equations (ODEs).

The time of a composite model $ C $ progresses when all the signal statuses $ s_{st} $ are resolved from unknowns $ \bot $ to either 0 or 1. In the presence of the instantaneous cycle, not all signal statuses may be resolved that blocks the model simulation from progressing to the next tick. This paper tackles this problem via an online method to prevent such a deadlock in composite models being executed with the SR MoC where the ICD is performed at every tick boundary.

\subsection{Transitive Closure and Reduction}\label{sec:tc}

Given a directed acyclic graph (DAG) $ G = \langle V, E \rangle $ where $ V $ is a set of vertices and $ E \subseteq V \times V $, transitive closure (TC) of $ G $ denoted as $ R^+ $ is the smallest transitive relation on $ V $ that includes $ E $ as a subset. $ \forall (u,v) \in V \times V,\ (u,v) \in R^+ $ iff there exist a path from $ u $ to $ v $ in $ E $. Therefore transitive closure of $ G $ gives a reachability relation between all vertices in $ V $. One can build a TC between all input and output vertices in $ G $ via matrix multiplication of an adjacency matrix $ \mathbf{A} $ in $ O(n^\omega) $ time where $ \omega = 2.38 $~\cite{le2014powers}.

Dynamic transitive closure (DTC) maintains a data-structure that can be updated upon insertion and addition of edges in the DAG. There are three main operations for a DTC:

\begin{itemize}
\item \textit{insert(x,y)} -- To add a transitive relation in a DTC between two vertices from $ x $ to $ y $.
\item \textit{delete(x,y)} -- To remove a transitive relation in a DTC between two vertices from $ x $ to $ y $.
\item \textit{query(x,y)} -- To check if a vertex $ y $ is \textit{reachable} from $ x $.
\end{itemize}
Complexities of these operations vary depending on the implementation of the data-structure. Typically, improving performance of one operation degrades the others and vice-versa~\cite{kapron2013dynamic,demetrescu2005trade}.

Transitive reduction (TR) of $ G $ denoted as $ R^- $ is a minimal set of $ E $ whose transitive closure is identical to the transitive closure of $ G $. In other words, it gives a graph that has the minimal number of edges with the same reachability relation as that of the original DAG. It is known that the time complexity of computing TR is in the same class as that of computing TC~\cite{aho1972transitive}. As it will be shown in the later section, we employ both TC and TR for reducing the stall time of the ICD.

\section{The Oracle-based Online Instantaneous Cycle Detection}\label{sec:met}

To enable the analysis of grey-box component-based models, we assume input-to-output dependencies are embedded in each model by the modelling tools such as in Functional Mock-up Unit \cite{gomes2018co}. In addition, we also assume the models can provide the updated dependency information upon the change of their modes. 
Since IP-related issue is one of the main reasons that the model vendors might hide the implementation of their model's behaviour, our method utilises a single trusted entity so-called an oracle. It thus limits the amount of information that must be (i) exchanged among model vendors, or (ii) provided to the models integrator, for the ICD.

\subsection{An Overview of the Analysis}\label{sec:overview}

\begin{figure}[t!]
\centering
\includegraphics[width=0.75\linewidth,page=4]{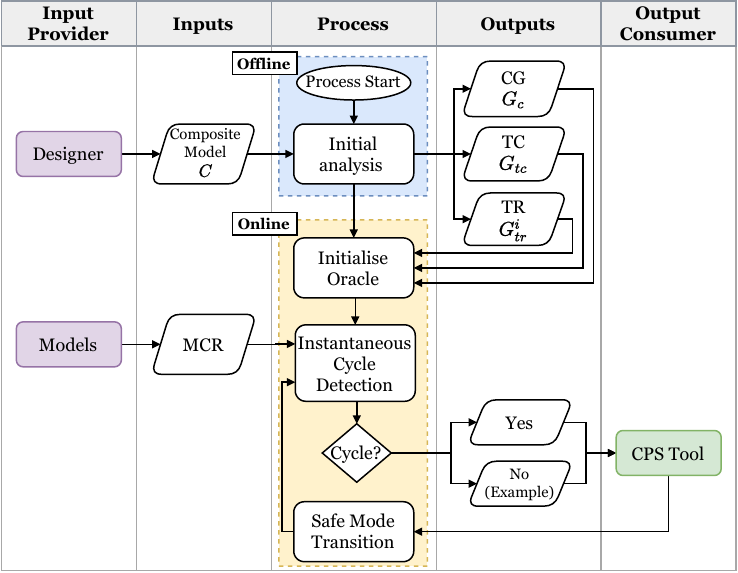}
\caption{Workflow diagram for the oracle-based instantaneous cycle detection}
\label{fig:wf}
\end{figure}

The overall workflow of the ICD is shown in Fig.~\ref{fig:wf}. The process is divided into two phases: offline and online, which are indicated by the blue and yellow boxes, respectively. At the beginning of the offline process, a composite model comprised of a set of interconnected sub-components via input and output ports is given as an input by a designer. This model is processed during the initial analysis phase, which generates three types of data structures. The first two are a \textit{composite graph} $ G_c $ and its transitive closure indicated by $ G_{tc} $. The other is a set of graphs $ \{G^1_{tr},...,G^k_{tr} \}$ for some $ k $ where each $ G^i_{tr} $ is obtained from the transitive reduction on the vertex-partitioned sub-graphs of $ G_{tc} $. A brief discussion on how to partition $ G_{tc} $ based on MCR types is presented in Section~\ref{sec:on}. 

\begin{figure}[t!]
\centering
\includegraphics[width=0.46\linewidth,page=5]{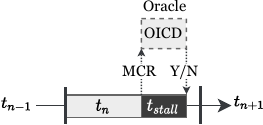}
\caption{An effect of the stall time for model simulation}
\label{fig:exec}
\end{figure}

The outputs of the offline phase are given as inputs to the oracle at the beginning of the online phase as indicated by the yellow box in Fig.~\ref{fig:wf}. When models need to change their modes, the information on their input to output dependency is transmitted to the oracle via an MCR message. The oracle chooses one of $ G_c $, $ G_{tc} $ and $ G^i_{tr} $ to perform the ICD and answers either `yes' or `no' back to the models. The time between the MCR and the answer generated by the oracle is called \textit{stall time}, during which all the models are blocked and cannot advance to the next simulation step. This is illustrated in Fig.~\ref{fig:exec} where the MCR call is made to the oracle at the end of the current synchronous tick $ n $ and the subsequent execution is blocked for the duration of $ t_{stall} $ due to online ICD (OICD).

\subsection{Graph Construction during the Offline Analysis}\label{sec:offline}

The construction of $ G_c = \langle V_c, E_c \rangle $ from the composite model $ C $ involves creating vertices $ V_c = I \cup O $ and an edge set $ E_c = S \cup E_L $ where $ E_L $ is a set of input to output edges for the initial modes for all models $ m \in M $ and $ S $ is a set of interface signals as explained in Section~\ref{ssec:sr}. $ S $ is fixed throughout the life-time of $ C $ whereas $ E_L $ changes along with the internal modes of the models within $ C $. The presence of cycles in this graph is considered as the presence of instantaneous cycles in the composite model:
\begin{definition}[Instantaneous cycle]
A composite model $ C $ contains the instantaneous cycle if its composite graph $ G_c $ contains any cycle.
\end{definition}
Then the problem is simply to check the presence of cycle directly on $ G_c $ using depth-first search (DFS) whenever the structure of the graph changes due to the mode changes in the models. However, as shown in \cite{zhou2008causality}, the complexity of DFS for the whole $ G_c $ would be costly especially with the large numbers of nodes and edges in the graph. Therefore we construct a transitive closure $ G_{tc} $ of $ G_c $ during the offline phase shown in Fig.~\ref{fig:wf} that can be used for efficiently querying the existence of instantaneous cycle in the composite model upon a mode change. 

Formally, $ G_{tc} = \langle V_{c}, E_{tc} \rangle $ is a transitive closure of $ G_c $. In this work, we adopt the technique introduced in~\cite{king2002fully} to compute $ G_{tc} $, which maintains the adjacency matrix explicitly. The adjacency matrix $ \mathbf{A}_{xy} $ of $ G_{tc} $ contains the number of paths for all $ (x,y) \in E_{tc} $; the existence of a path between two vertices in $ G_{tc} $ can be checked in $ O(1) $ time. Maintaining the number of paths in $ \mathbf{A} $ allows us to dynamically update $ G_{tc} $ upon addition or deletion of $ (x,y) \in E_{c} $ using the following algorithm~\cite{king2002fully}:
\begin{equation}\label{eq:tcupate}
\forall u \in pred(x),\ \forall v \in succ(y),\ \mathbf{A}_{uv} \gets \mathbf{A}_{uv} \pm \mathbf{A}_{ux}\cdot \mathbf{A}_{yv} 
\end{equation}
where $ pred(x) $ is a set of all direct predecessors of $ x $ and $ succ(y) $ is a set of all direct successors of $ y $. The $ \pm $ sign is $ + $ for insert and $ - $ for delete operation. The instantaneous cycle can be checked before $ (x,y) $ is inserted to $ E_c $ (and equivalently $ E_{tc} $) in $ O(1) $ by the following query:
\begin{equation}\label{eq:query}
x \in succ(y)
\end{equation}
For a sparse graph, the runtime complexity of Eq.~\eqref{eq:tcupate} is $ O(|V_c|^2) $. For a single edge update, Eq.~\eqref{eq:query} gives the fastest stall time. In case when there are multiple edges to be updated in a single MCR, the oracle has to perform Eq.~\eqref{eq:tcupate} as many times as the number of edges that are inserted if it only maintains $ G_{tc} $. An alternative method would be inserting or removing all edges requested by the MCR in $ G_c $ all at once and perform DFS for checking an instantaneous cycle. Again this would not be amenable if the size of $ G_c $ is large. 

To alleviate the time required to check instantaneous cycles in a large $ G_c $ and to handle multiple edge updates in a single MCR, we perform transitive reduction on vertex-partitioned sub-graphs $ G^i_{tc}=\langle V^i_{c},E^i_{tc}\rangle  $ of $ G_{tc} $ where $\forall_{0<i\leq k} V^i_{c} \subseteq V_{c}$, $ \bigcup_{i=1}^{k} V^i_{c} =V_{c} $, $\forall_{0<i,j\leq k,i\neq j} V^i_{c} \cap V^j_{c} = \emptyset $ and $ E^i_{tc} = \{(u,v) \mid \forall u,v \in V^i_{c}, (u,v) \in E_{tc}\} $. The idea is to create a set of $ G^i_{tr}=\langle V^i_{c},E^i_{tr}\rangle$ for $ 0<i\leq k $ where each $ G^i_{tr} $ is obtained from the transitive reduction of $ G^i_{tc} $ as explained in Section~\ref{sec:tc}. Since $ G^i_{tr} $ has the same reachability relation as $ G^i_{tc} $ (which is the same as that of $ G_{tc} $ and $ G_c $ for the vertex partition $ V_c^i $) but with the minimal number of edges, the instantaneous cycle detection on $ G^i_{tr} $ via DFS for some $ i $ would be faster than on the entire $ G_c $.

\subsection{Online Analysis}\label{sec:on}

The oracle adaptively chooses one of $ G_c $, $ G_{tc} $ and $ G^i_{tr} $ to perform the ICD. Formally, an MCR call made by a model is a sequence of edges $ E_{mcr}=\{e^l_n\}_{i=1}^j = \{e^l_1,\dots,e^l_j \} $ to be updated in $ G_c $ where $ l $ indicates insertion (1) and deletion (0) of an edge. A response to the model $ \gamma \in \{1, 0\} $ from the oracle indicates either an acceptance (1) or rejection (0) for the corresponding MCR. Next, we define three types of MCR calls:

\begin{enumerate}
\item \textit{MCR type a} -- When a single model makes an MCR with a change in a single input to output dependency, i.e. $ |E_{mcr}| = 1 $.

\item \textit{MCR type b} -- When an $ E_{mcr} $ consists of changes in multiple input to output dependencies where \mbox{$ \exists_{=1} G^i_{tr} \in \{G^1_{tr},...,G^k_{tr}\} $} (there exists a \textit{unique} $ G^i_{tr} $ in the set) such that $ \forall (x,y) \in E_{mcr} ,\ x,y \in V^i_{c}$.

\item \textit{MCR type c} --  When an MCR consists of changes in multiple input to output dependencies where $ \exists G_{tr}^i,G_{tr}^j \in \{G^1_{tr},...,G^k_{tr}\},\ i\neq j,\ E_{mcr} \cap E_{tr}^i \neq \emptyset \wedge E_{mcr} \cap E_{tr}^j \neq \emptyset $.
\end{enumerate}

\begin{figure}[t!]
\footnotesize
\begin{algorithmic}[1]
\Function{DetectCycle}{$ G_c,G_{tc}, \{G^1_{tr},...,G^k_{tr}\}, E_{mcr}$}
\If{$ |E_{mcr}| = 1$} \Comment{MCR type-a} \label{al:1}
\State Perform a cycle check using Eq.~\eqref{eq:query}. \label{al:2}
\ElsIf{$ \exists_{=1} G^i_{tr} \text{ s.t. } \forall (x,y) \in E_{mcr} ,\ x,y \in V^i_{c}  $} \label{al:b}
\State Update $ G^i_{tr} $ with $ E_{mcr} $ \Comment{MCR type-b} \label{al:up1}
\State Perform DFS on $ G^i_{tr} $ for presence of any cycles. \label{al:3}
\Else \Comment{MCR type-c} \label{al:c}
\State Update $ G_c $ with $ E_{mcr} $ \label{al:up2}
\State Perform DFS on $ G_c $ for presence of any cycles. \label{al:4}
\EndIf \label{al:5}
\State Send results to models
\If{Instantaneous cycle is found}
\State Revert updates on $ G_{tr}^i $ or $ G_c $ \label{al:rev}
\Else
\State \Call{UpdateGraphs}{$ G_c,G_{tc}, \{G^1_{tr},...,G^k_{tr}\}, E_{mcr} $} \label{al:up3}
\EndIf
\EndFunction
\end{algorithmic}
\caption{Pseudocode for the online ICD}
\label{alg:cyc}
\end{figure}

The decision on how to divide $ G_{tc} $ into a set of sub-graphs for computing $ G^i_{tr} $ depends on applications and the analysis on the frequency of edge updates in the MCRs. For example, we can group vertices in the same $ G^i_{tr} $ when all $ v \in V^i_{c} $ appear frequently in the MCRs in the same synchronous tick. Another possibility is to consider each $ G^i_{tr} $ belongs to a single model $ m \in M $ since all $ v \in V^i_{c} $ would most likely appear in the same synchronous tick. In this way, the oracle will receive more MCRs of type $ b $, which are faster to analyse than MCRs of type $ c $. It should be noted that upon mode change, models only have to provide changes in their input to output dependencies in $ E_{mcr} $ to minimise the amount of data need to be transferred to the oracle. Furthermore, oracle does not need to consider all possible combinations of modes from sub-models since only dependency changes in the target modes are needed for the analysis.

The pseudocode of the oracle that performs the online ICD is shown in Fig.~\ref{alg:cyc}. The MCR type $ a $ is checked using Eq.~\eqref{eq:query} as shown at lines~\ref{al:1}-\ref{al:2}. For the MCR type $ b $ (line~\ref{al:b}), we check if there exist a pre-computed $ G^i_{tr} $ where all the edge updates required by the MCR affect vertices in $ V_c^i $ are in $ E^i_{tr} $. If it is the case, a DFS is performed on $ G^i_{tr} $ for ICD, which is shown at line~\ref{al:3}. For the MCR type $ c $ (line~\ref{al:c}), the edges in $ E_{mcr} $ are across multiple $ G^i_{tr} $. In this case, the oracle performs DFS on $ G_c $ as a fallback option as shown at line~\ref{al:4}. Note that both $ G^i_{tr} $ and $ G_c $ are updated with $ E_{mcr} $ before the analysis as shown at lines~\ref{al:up1} and \ref{al:up2}. Upon detection of an instantaneous cycle, these updates are reverted at line~\ref{al:rev}.

\begin{figure}[t!]
\footnotesize
\algrenewcommand\algorithmicindent{1.0em}%
\begin{algorithmic}[1]
\Function{UpdateGraphs}{$ G_c,G_{tc}, \{G^1_{tr},...,G^k_{tr}\}, E_{mcr}$}
\If{$ |E_{mcr}| > c $} \label{alg:ft}
\State \Call{RecomputeTC}{$ G_{c} $} \label{alg:2}
\State \Call{UpdateTR}{$ G_c, G_{tc}, \{G^1_{tr},...,G^k_{tr}\} $} \label{alg:3}
\State \textbf{return}
\EndIf \label{alg:reee}

\State $ \mathbf{A} \gets $ adjacency matrix of $ G_{tc} $
\For{$ \forall_{i \in \{1,\dots,|E_{mcr}| \}} (x,y)_i \in E_{mcr} $}
\ForAll{$ u \in pred(x) $} \label{alg:tc1}
\ForAll{$ v \in succ(y) $}
\State $\mathbf{A}_{uv} \gets \mathbf{A}_{uv} \pm \mathbf{A}_{ux}\cdot \mathbf{A}_{yv} $ \label{alg:tcc}

\EndFor
\EndFor \label{alg:tc2}
\EndFor
\State \Call{UpdateTR}{$ G_c, G_{tc}, \{G^1_{tr},...,G^k_{tr}\}, E_{mcr} $} \label{alg:5}
\EndFunction
\end{algorithmic}
\caption{Pseudocode for updating $ G_{tc} $}
\label{fig:utc}
\end{figure}

After the ICD is done, the oracle updates all three graphs which is indicated by the function call \textsc{UpdateGraphs} at line~\ref{al:up3} in Fig.~\ref{alg:cyc}. The pseudocode of \textsc{UpdateGraphs} is shown in Fig.~\ref{fig:utc} where the lines~\ref{alg:tc1}-\ref{alg:tcc} implements Eq.~\eqref{eq:tcupate} that updates $ G_{tc} $. During the execution of \textsc{UpdateGraphs}, the oracle may receive a new MCR and we assume it can preempt the execution of \textsc{UpdateGraphs} and perform ICD for the new MCR. However, the oracle cannot use $ G_{tc} $ or $ G^i_{tr} $ for the analysis for the new request because these graphs are not up-to-date with the current modes of the models. Instead, DFS on $ G_c $ is performed as a fallback method, which algorithm is identical to the lines~\ref{al:up2}-\ref{al:4} in Fig.~\ref{alg:cyc}. In this case, we also assume that the previous invocation of \textsc{UpdateGraphs} is abandoned and it is invoked again after checking instantaneous cycle. The new invocation would include all edge updates from the previous MCRs that have not yet been processed by \textsc{UpdateGraphs}.

The size of $ E_{mcr} $ in \textsc{UpdateGraphs} can grow unbounded if the inter-arrival time of MCRs is faster than the speed of the update process of \mbox{\textsc{UpdateGraphs}}. Therefore, \textsc{UpdateGraphs} invokes \textsc{RecomputeTC} if the size of accumulated $ E_{mcr} $ is bigger than certain value $ c $, which is shown at lines~\ref{alg:ft}-\ref{alg:reee}. In other words, if the time required to update $ G_{tc} $ from $ E_{mcr} $ becomes longer than recomputing $ G_{tc} $ from the scratch, the algorithm invokes \textsc{RecomputeTC}.

\begin{figure}[t!]
\footnotesize
\algrenewcommand\algorithmicindent{1.0em}%
\begin{algorithmic}[1]
\Function{UpdateTR}{$ G_c, G_{tc}, \{G^1_{tr},...,G^k_{tr}\}, E_{mcr} $}
\For{$ i \in \{1,...,k\} $} \label{l:l-s}
\State $ \mathbf{A} \gets $ adjacency matrix for $ G^i_c \subseteq G_c $
\State $ \forall_{u,v \in \{1,\cdots,|V^i_{c}|\}}\ \mathbf{B}_{uv} \gets \mathtt{query}_{tc}(u,v) $ \label{al:bq}
\State $ \mathbf{C} \gets \mathbf{A} \cdot \mathbf{B} $  \label{alg:mm}
\State $ \forall_{u,v \in \{1,\cdots,|V^i_{c}|\}}\ \mathbf{C}_{uv} \gets 1\ \mathtt{if}\ \mathbf{A}_{uv} > 0 \wedge \mathbf{C}_{uv} = 0\ \mathtt{else}\ 0 $ \label{alg:adjset}
\State Set adjacency matrix for $ G^i_{tr} $ to $ \mathbf{C} $ \label{alg:setc}
\EndFor \label{l:l-end}
\EndFunction
\end{algorithmic}
\caption{Pseudocode for updating all $ G^i_{tr} \in \{G^1_{tr},...,G^k_{tr}\} $}
\label{fig:utr}
\end{figure}

All $ G^i_{tr} $ are updated in \textsc{UpdateTR} whose pseudocode is shown in Fig.~\ref{fig:utr}. The main part of this algorithm, based on the method introduced in \cite{aho1972transitive}, is the multiplication of the adjacency matrices $ \mathbf{A} $ and $ \mathbf{B} $ at line~\ref{alg:mm} for $ G^i_c $ and $ G^i_{tc} $ where $ G^i_c = \langle V^i_c,E^i_c \rangle $, $ E^i_c = \{(u,v) \mid \forall u,v \in V^i_{c}, (u,v) \in E_{c}\} $. It should be noted that $ G^i_{tc} $ is \textit{not} a transitive closure of $ G^i_c $. If the adjacency matrix of $ G^i_{tc} $ is not maintained explicitly, the algorithm has to make queries on $ G_{tc} $ to create $ \mathbf{B} $ as shown at line~\ref{al:bq}. The result of the multiplication $ \mathbf{C} $ is further processed at line~\ref{alg:adjset} where each element at $ \mathbf{C}_{ij} $ is set to 1 if its previous value was 0 and if the element in $ \mathbf{A}_{ij} $ at the same index is greater than 0. Otherwise, the value at $ \mathbf{C}_{ij} $ is set to 0. The adjacency matrix for $ G_{tr}^i $ is then updated with $ \mathbf{C} $ (line~\ref{alg:setc}).

\subsection{Complexity and Correctness of the Online Analysis}\label{sec:complexity}

The complexity and correctness of the online ICD is presented in this section. The update time $ \Delta $ of a \textit{single} $ G^i_{tr} $ by the oracle is 
\begin{equation}
\Delta = \delta_{q} + \delta_{tr}
\end{equation}
where $ \delta_{q} $ is the time required to make $ |V^i_{c}|^2 $ reachability queries on $ G_{tc} $ and $ \delta_{tr} $ is the actual time required to compute $ G^i_{tr} $. It is shown in~\cite{aho1972transitive} that the time complexity to compute transitive reduction is in the same class as the transitive closure computation. Therefore, the fastest known algorithm to compute $ G^i_{tr} $ is boolean matrix multiplication, and hence  $O(\delta_{tr})= O(|V^i_{c}|^\omega) $. To compute $ \delta_q $ we need to make reachability queries on $ G_{tc} $ as many as $ |V^i_{c}|^2 $. Yet if we maintain $ \mathbf{A} $ of $ G_{tc} $ explicitly, $ \delta_q = 0 $. Therefore $ \Delta = O(|V^i_{c}|^\omega) $. Then, runtime complexities of \textsc{RecomputeTC} and \textsc{UpdateTR} are presented in the following lemma:
\begin{lemma}\label{lem:1}
The worst-case runtime complexity of the functions \textsc{RecomputeTC} and \textsc{UpdateTR} is $ O(|V_c|^\omega) $ where $ \omega = 2.38 $ is the greatest lower bound for the matrix multiplication.
\end{lemma}
\begin{proof}
The runtime complexity of transitive closure (\textsc{RecomputeTC}) is known to be in the same class as that of matrix multiplication, which is $ O(|V_c|^\omega) $. The runtime complexity of \textsc{UpdateTR} is $ O(\sum_{i=1}^k|V^i_c|^\omega) $ where $ k $ is the total number of vertex-partitioned sub-graphs $ G^i_{tc} $ and $ |V^i_c|^\omega $ is from the matrix multiplication as shown at lines~\ref{l:l-s} and \ref{alg:mm} in Fig.~\ref{fig:utr}, respectively. From the multinomial and generalised binomial theorems, we can see $ |V_c|^\omega > \sum_{i=1}^k|V^i_c|^\omega $ where $ |V_c| = \sum_{i=1}^k|V^i_c| $ for any $ i > 1 $ and $ \omega > 1$. Therefore, the complexity of \textsc{UpdateTR} is reduced to $ O(|V_c|^\omega) $.
\end{proof}

It is obvious that, as shown at lines~\ref{al:1}-\ref{al:5} in Fig.~\ref{alg:cyc}, the worst-case time bound for the stall time of the models when the oracle was in \textsc{DetectCycle} is dominated by the DFS on $ G_c $, which is $ O(|V_c|^2) $, for the MCR type $ c $. Since the oracle is able to preempt the graph update process in \textsc{UpdateGraphs} upon receiving new MCRs from the models, the worst-case stall time is also $ O(|V_c|^2) $. However, note the physical stall times are expected to be much lower than this worst-case, especially if the oracle is able to frequently exploit the benefits of MCR types $ a $ and $ b $. Time complexity to check if the incoming MCR is type $ b $ (line~\ref{al:b} in Fig.~\ref{alg:cyc}) is $ O(|E_{mcr}|) $ where $ |E_{mcr}| \leq |E_c| $ since for each $ (x,y) \in E_{mcr} $, the oracle can check in which vertex-partitioned $ V^i_c $ $ x $ and $ y $ are mapped in $ O(1) $.

The following theorem proves the correctness of our ICD technique:

\begin{theorem}
The algorithm in Fig.~\ref{alg:cyc} correctly detects instantaneous cycles for all three MCR types.
\end{theorem}
\begin{proof}
By definition, $ G_{tc} $ has reachability relation between any vertices in $ V_c $ and the relation is transitive. Then for the MCR type $ a $, it is trivial to check if an edge $ (x,y) \in E_{mcr} $ would create a cycle in $ G_c $ via the query shown in \eqref{eq:query} on $ G_{tc} $. Concretely, $ G_{tc} $ contains the transitive relation of $ (x,y) $ in $ G_c $ where $ x \in succ(y) $ iff there exist a path from $ y $ to $ x $ in $ G_c $. Therefore, one can verify if the addition of an edge $ (x,y) $ would create a cycle via checking the existence of a path from $ y $ to $ x $ via $ x \in succ(y) $. For the MCR type $ b $, since any vertex-partitioned sub-graph $ G^i_{tc} $ has a set of edges $ E^i_{tc} = \{(u,v) \mid \forall u,v \in V^i_{c}, (u,v) \in E_{tc}\} $, $ G^i_{tc} $ also has the same reachability relation between any vertices in $ V^i_c $ as $ G_{tc} $ and $ G_c $ and is transitive. By definition, transitive reduction of $ G^i_{tc} $ denoted as $ G^i_{tr} $ has the same reachability relation as $ G^i_{tc} $ and $ G_c $ for the vertex set $ V_c^i $. Therefore, for any given $ G_c $ and $ G_{tr}^i $, if there exists a set of edges $ E_{cycle} \subseteq E_c $ that creates a cycle in $ G_c $, then there also exists the same set of edges that creates a cycle in $ G_{tr}^i $ such that $ E_{cycle} \subseteq E_{tr}^i $ and vice-versa. For the MCR type $ c $, it is trivial to see any DFS algorithm can find the existence of a cycle in $ G_c $, and this concludes the proof.
\end{proof}

\section{Experiment: Finding the Saturation Point from the Online Instantaneous Cycle Detection}\label{sec:exp}

In this section, we present a set of experiments using synthetically generated examples where the oracle performs the online ICD based on incoming MCR types. 
All of the benchmarks were written in MATLAB version 2019b and carried out on the machine with a Core i7-8650U CPU @ 1.9Ghz laptop with 16GB RAM. 

\subsection{Experimental Settings}

\begin{table}[t!]
\centering
\caption{$ G_c $ candidates for the experiment}
\label{tab:1}
\begin{tabular}{d{0.1}|d{0.10}|d{0.10}|d{0.12}|d{0.12}|d{0.12}}
\hline
$ G_{c} $ & $ |V_c| $ & $ |M| $ & $ |E| $ & $ |M_{mcr}| $ & $ |E_{mcr}| $ \\
\hline\hline
$ G_{c,100} $ & 100 & 5 & 1274 & 3 & 6 \\
$ G_{c,200} $ & 200 & 10 & 4943 & 10 & 20 \\
$ G_{c,400} $ & 400 & 10 & 19991 & 10 & 20 \\
$ G_{c,600} $ & 600 & 10 & 44880 & 10 & 30 \\
$ G_{c,800} $ & 800 & 10 & 79900 & 10 & 30 \\
\hline
\end{tabular}
\end{table}

The main objective of this experiment is to investigate how fast the models can make MCR before the oracle always falls back to DFS on $ G_c $ for ICD. We call this threshold a \textit{saturation point}. Table~\ref{tab:1} shows an overview of the experimental candidates for $ G_c $. We have generated a random set of $ G_c $ in MATLAB based on the parameters shown in the table. The size of ports (vertices) in $ G_c $ is increased up to 800 to visualise the trend of the saturation points for different sizes of $ G_c $. As mentioned in Section~\ref{sec:offline}, we assume each $ G^i_{tr} $ is a transitive reduction of $ G^i_{tc} $ for a model $ m_i \in M $. The ports in $ V_c $ are evenly distributed among the models and the ratio of input to output ports in $ V^i_c $ for each model is set to 1. We use the notation $ G_{c,n} $ for indicating a composite graph $ G_c $ with $ |V_c|=n $. The number of models $ |M| $ is fixed to 10 for all candidates except $ G_{c,100} $, which has 5. This is to increase the complexity of the ICD for MCR type $ b $ along with the $ G_c $ size, i.e. to increase the size of each $ G^i_{tr} $. The edges are created based on the uniformly distributed random numbers such that $\forall_{u,v\in V}\ P((u,v) \in E) = 0.5 $. $ |M_{mcr}| $ is the maximum number of models and $ |E_{mcr}| $ is the number of edges to update per MCR, respectively. The types of MCR calls made by the models are evenly distributed in this experiment for each $ a $, $ b $ and $ c $ and in total 100 calls are made altogether. Finally, we designed this experiment in a way that the models only send MCR calls that do not create instantaneous cycles. This results in the longest possible stall time for each MCR call since the oracle required to search the complete vertices and edges in the graph.

\begin{figure*}[t!]
\centering
\subfloat[$ G_{c,100} $]{
\begin{tikzpicture}[baseline]
\pgfplotstableread[col sep=comma,]{results/Period-100-5.csv}\tabledata
\pgfplotsset{cycle list/Greys-3}
\begin{axis}[
	ytick style={draw=none},
	table/col sep = comma, 
	mark options={scale=0.5}, 
	ultra thin,
	scaled x ticks=true,
	width=0.5\linewidth,
	ytick={1,2},
	yticklabels from table={\tabledata}{y},
	height=3.5cm,
	xmin=0,
	ylabel=$ P_{mcr} $ (s),
	xlabel=Stall time (s),
	],
\addplot+ [boxplot, fill, draw=black] table [y=0.001]{\tabledata};
\addplot+ [boxplot, fill, draw=black] table [y=0.002]{\tabledata};
\end{axis}
\end{tikzpicture}
\label{fig:r1}
}
\subfloat[$ G_{c,200} $]{
\begin{tikzpicture}[baseline]
\pgfplotstableread[col sep=comma,]{results/Period-200-10.csv}\tabledata
\pgfplotsset{cycle list/Greys-6}
\begin{axis}[
	ytick style={draw=none},
	table/col sep = comma, 
	mark options={scale=0.5}, 
	ultra thin,
	scaled x ticks=true,
	width=0.5\linewidth,
	ytick={1,2,3,4,5,6},
	yticklabels from table={\tabledata}{y},
	height=3.5cm,
	xmin=0,
	ylabel=$ P_{mcr} $ (s),
	xlabel=Stall time (s),
	],
\addplot+ [boxplot, fill, draw=black] table [y=0.005]{\tabledata};
\addplot+ [boxplot, fill, draw=black] table [y=0.01]{\tabledata};
\addplot+ [boxplot, fill, draw=black] table [y=0.015]{\tabledata};
\addplot+ [boxplot, fill, draw=black] table [y=0.02]{\tabledata};
\addplot+ [boxplot, fill, draw=black] table [y=0.025]{\tabledata};
\addplot+ [boxplot, fill, draw=black] table [y=0.03]{\tabledata};
\end{axis}
\end{tikzpicture}
\label{fig:r2}
}

\subfloat[$ G_{c,400} $]{
\begin{tikzpicture}[baseline]
\pgfplotstableread[col sep=comma,]{results/Period-400-10.csv}\tabledata
\pgfplotsset{cycle list/Greys-6}
\begin{axis}[
	ytick style={draw=none},
	table/col sep = comma, 
	mark options={scale=0.5}, 
	ultra thin,
	scaled x ticks=true,
	width=0.5\linewidth,
	ytick={1,2,3,4,5,6},
	yticklabels from table={\tabledata}{y},
	height=3.5cm,
	xmin=0,
	ylabel=$ P_{mcr} $ (s),
	xlabel=Stall time (s),
	],
\addplot+ [boxplot, fill, draw=black] table [y=0.13]{\tabledata};
\addplot+ [boxplot, fill, draw=black] table [y=0.14]{\tabledata};
\addplot+ [boxplot, fill, draw=black] table [y=0.15]{\tabledata};
\addplot+ [boxplot, fill, draw=black] table [y=0.16]{\tabledata};
\addplot+ [boxplot, fill, draw=black] table [y=0.17]{\tabledata};
\addplot+ [boxplot, fill, draw=black] table [y=0.18]{\tabledata};
\end{axis}
\end{tikzpicture}
\label{fig:r3}
}%
\subfloat[$ G_{c,600} $]{
\begin{tikzpicture}[baseline]
\pgfplotstableread[col sep=comma,]{results/Period-600-10.csv}\tabledata
\pgfplotsset{cycle list/Greys-6}
\begin{axis}[
	ytick style={draw=none},
	table/col sep = comma, 
	mark options={scale=0.5}, 
	ultra thin,
	scaled x ticks=true,
	width=0.5\linewidth,
	ytick={1,2,3,4,5,6,7},
	yticklabels from table={\tabledata}{y},
	height=3.5cm,
	xmin=0,
	ylabel=$ P_{mcr} $ (s),
	xlabel=Stall time (s),
	],
\addplot+ [boxplot, fill, draw=black] table [y=0.6]{\tabledata};
\addplot+ [boxplot, fill, draw=black] table [y=0.625]{\tabledata};
\addplot+ [boxplot, fill, draw=black] table [y=0.65]{\tabledata};
\addplot+ [boxplot, fill, draw=black] table [y=0.675]{\tabledata};
\addplot+ [boxplot, fill, draw=black] table [y=0.7]{\tabledata};
\addplot+ [boxplot, fill, draw=black] table [y=0.725]{\tabledata};
\addplot+ [boxplot, fill, draw=black] table [y=0.75]{\tabledata};
\end{axis}
\end{tikzpicture}
\label{fig:r4}
}

\subfloat[$ G_{c,800} $]{
\begin{tikzpicture}[baseline]
\pgfplotstableread[col sep=comma,]{results/Period-800-10.csv}\tabledata
\pgfplotsset{cycle list/Greys-6}
\begin{axis}[
	ytick style={draw=none},
	table/col sep = comma, 
	mark options={scale=0.5}, 
	ultra thin,
	scaled x ticks={base 10:2},
	width=0.5\linewidth,
	ytick={1,2,3,4,5,6},
	yticklabels from table={\tabledata}{y},
	height=3.5cm,
	xmin=0,
	ylabel=$ P_{mcr} $ (s),
	xlabel=Stall time (s),
	],
\addplot+ [boxplot, fill, draw=black] table [y=1.5]{\tabledata};
\addplot+ [boxplot, fill, draw=black] table [y=1.6]{\tabledata};
\addplot+ [boxplot, fill, draw=black] table [y=1.7]{\tabledata};
\addplot+ [boxplot, fill, draw=black] table [y=1.8]{\tabledata};
\addplot+ [boxplot, fill, draw=black] table [y=1.9]{\tabledata};
\addplot+ [boxplot, fill, draw=black] table [y=2]{\tabledata};
\end{axis}
\end{tikzpicture}
\label{fig:r5}
}
\caption{Distributions of stall times for \protect\subref{fig:r1} $ G_{c,100} $, \protect\subref{fig:r2} $ G_{c,200} $, \protect\subref{fig:r3} $ G_{c,400} $, \protect\subref{fig:r4} $ G_{c,600} $ and \protect\subref{fig:r5} $ G_{c,800} $}
\label{fig:results1}
\end{figure*}

\subsection{Benchmark Results}\label{sec:b-result}

The stall time distributions for each experiment are shown in Fig.~\ref{fig:results1} with varying MCR call periods denoted as $ P_{mcr} $. The smallest $ P_{mcr} $ on the y-axis in each figure indicates the saturation point for the oracle whereas the largest $ P_{mcr} $ is when all of the MCRs are served without falling back to DFS on $ G_c $. We say the oracle is saturated when the oracle starts to perform ICD for all MCRs using the fallback method, i.e. all MCRs are treated as type $ c $.

For $ G_{c,100} $ shown in Fig.~\ref{fig:r1}, the oracle could serve MCRs without saturation up to $ P_{mcr} = 0.001 $ whereas all of the MCRs are served without the fallback method when $ P_{mcr} > 0.002 $. For $ G_{c,200} $ shown in Fig.~\ref{fig:r2}, the interquartile range (IQR) of the box with $ P_{mcr}=0.005 $ is relatively shorter than other boxes. This is due to the small variations between the measured stall times since most of the MCR calls served by the oracle are done via the fallback method (DFS) on $ G_c $. On the other hand, the IQR becomes larger with longer $ P_{mcr} $ because the oracle completes updating the graphs more often before the arrival of the next MCR, and therefore more MCRs with types $ a $ and $ b $ are served by the oracle without the fallback method. The same trend can be seen in the experiments for $ G_{c,400} $, $ G_{c,600} $ and $ G_{c,800} $, which are shown in Figs.~\ref{fig:r3}, \ref{fig:r4} and \ref{fig:r5}, respectively. For $ G_{c,400} $, $ P_{mcr} $ had to be increased up to 0.18 second to avoid saturation whereas for $ G_{c,600} $ and $ G_{c,800} $ this had to be further increased up to 0.75 and 2 seconds, respectively. Note that some of the outliers shown in the figures are due to the cold start of the benchmark program and the measurements are quickly settled down after a few simulation steps. 

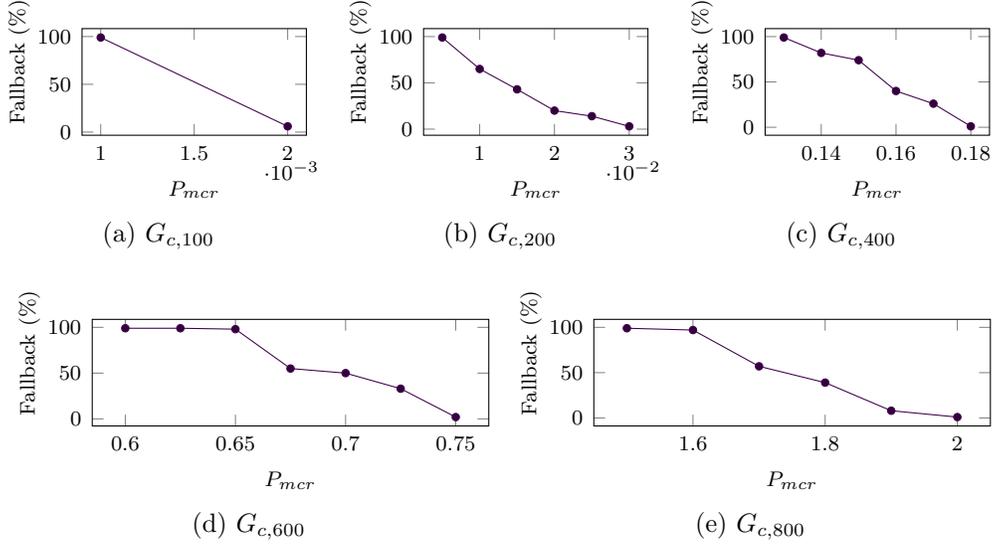
\begin{figure*}[t!]

\centering
\subfloat[$ G_{c,100} $]{
\begin{tikzpicture}[baseline]
\begin{axis}[
height=3.0cm,
width=0.333\linewidth,
ylabel style={font=\scriptsize},
xlabel style={font=\scriptsize},
ylabel=Fallback (\%),
xlabel=$ P_{mcr} $,
]
\addplot+ table[y=y, x=x] {results/Saturation-100-5.csv};
\end{axis}
\end{tikzpicture}
\label{fig:r2-1}
}
\subfloat[$ G_{c,200} $]{
\begin{tikzpicture}[baseline]
\begin{axis}[
height=3.0cm,
width=0.333\linewidth,
ylabel style={font=\scriptsize},
xlabel style={font=\scriptsize},
ylabel=Fallback (\%),
xlabel=$ P_{mcr} $
]
\addplot+ table[y=y, x=x] {results/Saturation-200-10.csv};
\end{axis}
\end{tikzpicture}
\label{fig:r2-2}
}
\subfloat[$ G_{c,400} $]{
\begin{tikzpicture}[baseline]
\begin{axis}[
height=3.0cm,
width=0.333\linewidth,
ylabel style={font=\scriptsize},
xlabel style={font=\scriptsize},
ylabel=Fallback (\%),
xlabel=$ P_{mcr} $
]
\addplot+ table[y=y, x=x] {results/Saturation-400-10.csv};
\end{axis}
\end{tikzpicture}
\label{fig:r2-3}
}

\subfloat[$ G_{c,600} $]{
\begin{tikzpicture}[baseline]
\begin{axis}[
height=3.0cm,
width=0.5\linewidth,
ylabel style={font=\scriptsize},
xlabel style={font=\scriptsize},
ylabel=Fallback (\%),
xlabel=$ P_{mcr} $
]
\addplot+ table[y=y, x=x] {results/Saturation-600-10.csv};
\end{axis}
\end{tikzpicture}
\label{fig:r2-4}
}
\subfloat[$ G_{c,800} $]{
\begin{tikzpicture}[baseline]
\begin{axis}[
height=3.0cm,
width=0.5\linewidth,
ylabel style={font=\scriptsize},
xlabel style={font=\scriptsize},
ylabel=Fallback (\%),
xlabel=$ P_{mcr} $
]
\addplot+ table[y=y, x=x] {results/Saturation-800-10.csv};
\end{axis}
\end{tikzpicture}
\label{fig:r2-5}
}
\caption{Percentage of MCRs served by the oracle via the DFS fallback method on $ G_c $ for \protect\subref{fig:r2-1} $ G_{c,100} $, \protect\subref{fig:r2-2} $ G_{c,200} $, \protect\subref{fig:r2-3} $ G_{c,400} $, \protect\subref{fig:r2-4} $ G_{c,600} $ and \protect\subref{fig:r2-5} $ G_{c,800} $}
\label{fig:result2}
\end{figure*}

More details on the proportion of the MCR calls served by the oracle via the fallback method are shown in Fig.~\ref{fig:result2}. The $ P_{mcr} $ values on the x-axis in each figure correspond to the $ P_{mcr} $ values on the y-axis in Fig.~\ref{fig:results1} of the same subfigure label. The results indicate that the number of fallback methods taken by the oracle decreases almost linearly with increasing $ P_{mcr} $ values. One interesting outcome is that the variations of the stall times are significantly reduced when the amount of fallback methods drops to approximately 50\%. This can be seen by comparing the results shown in Figures~\ref{fig:result2} and \ref{fig:results1} where the lower quartile values are increased such that they are close to median values. This indicates that at this point most of the MCR types $ a $ and $ b $, which are the fastest to check, are served via the fallback methods.

\begin{figure*}[t!]
\centering
%
\subfloat[$ G_{c,100} $]{
\begin{tikzpicture}[baseline]
\begin{axis}[
height=3.0cm,
width=0.3\linewidth,
every x tick label/.append style={alias=XTick,inner xsep=0pt},
every x tick scale label/.style={at=(XTick.base east),anchor=base west},
ylabel=Accumulated\\stall time (s),
xlabel=$ P_{mcr} $,
ylabel style={align=left,font=\scriptsize},
xlabel style={font=\scriptsize},
ybar,
xtick=data,
enlarge x limits=1.1,
ymin=0
]
\addplot+ table[y=y, x=x] {results/AccStall-100-5.csv};
\end{axis}
\end{tikzpicture}
\label{fig:r3-1}
}
\subfloat[$ G_{c,200} $]{
\begin{tikzpicture}[baseline]
\begin{axis}[
height=3.0cm,
width=0.3\linewidth,
every x tick label/.append style={alias=XTick,inner xsep=0pt},
every x tick scale label/.style={at=(XTick.base), anchor=north east},
ylabel=Accumulated\\stall time (s),
xlabel=$ P_{mcr} $,
ylabel style={align=left,font=\scriptsize},
xlabel style={font=\scriptsize},
ybar,
xtick=data,
enlarge x limits=0.21,
ymin=0,
bar width=5pt,
xticklabel style={rotate=90},
]
\addplot+ table[y=y, x=x] {results/AccStall-200-10.csv};
\end{axis}
\end{tikzpicture}
\label{fig:r3-2}
}
\subfloat[$ G_{c,400} $]{
\begin{tikzpicture}[baseline]
\begin{axis}[
height=3.0cm,
width=0.3\linewidth,
every x tick label/.append style={alias=XTick,inner xsep=0pt},
every x tick scale label/.style={at=(XTick.base), anchor=north east},
ylabel=Accumulated\\stall time (s),
xlabel=$ P_{mcr} $,
ylabel style={align=left,font=\scriptsize},
xlabel style={font=\scriptsize},
ybar,
xtick=data,
enlarge x limits=0.21,
ymin=0,
bar width=5pt,
xticklabel style={rotate=90},
]
\addplot+ table[y=y, x=x] {results/AccStall-400-10.csv};
\end{axis}
\end{tikzpicture}
\label{fig:r3-3}
}

\subfloat[$ G_{c,600} $]{
\begin{tikzpicture}[baseline]
\begin{axis}[
height=3.0cm,
width=0.5\linewidth,
every x tick label/.append style={alias=XTick,inner xsep=0pt},
every x tick scale label/.style={at=(XTick.base), anchor=north east},
ylabel=Accumulated\\stall time (s),
xlabel=$ P_{mcr} $,
ylabel style={align=left,font=\scriptsize},
xlabel style={font=\scriptsize},
ybar,
xtick=data,
enlarge x limits=0.12,
ymin=0,
bar width=4pt,
xticklabel style={rotate=90},
]
\addplot+ table[y=y, x=x] {results/AccStall-600-10.csv};
\end{axis}
\end{tikzpicture}
\label{fig:r3-4}
}
\subfloat[$ G_{c,800} $]{
\begin{tikzpicture}[baseline]
\begin{axis}[
height=3.0cm,
width=0.5\linewidth,
every x tick label/.append style={alias=XTick,inner xsep=0pt},
every x tick scale label/.style={at=(XTick.base), anchor=north east},
ylabel=Accumulated\\stall time (s),
xlabel=$ P_{mcr} $,
ylabel style={align=left,font=\scriptsize},
xlabel style={font=\scriptsize},
ybar,
xtick=data,
enlarge x limits=0.25,
ymin=0,
bar width=5pt,
xticklabel style={rotate=90},
]
\addplot+ table[y=y, x=x] {results/AccStall-800-10.csv};
\end{axis}
\end{tikzpicture}
\label{fig:r3-5}
}
\caption{Accumulated stall times for \protect\subref{fig:r2-1} $ G_{c,100} $, \protect\subref{fig:r2-2} $ G_{c,200} $, \protect\subref{fig:r2-3} $ G_{c,400} $, \protect\subref{fig:r2-4} $ G_{c,600} $ and \protect\subref{fig:r2-5} $ G_{c,800} $ for 100 MCR calls to the oracle}
\label{fig:result3}
\end{figure*}
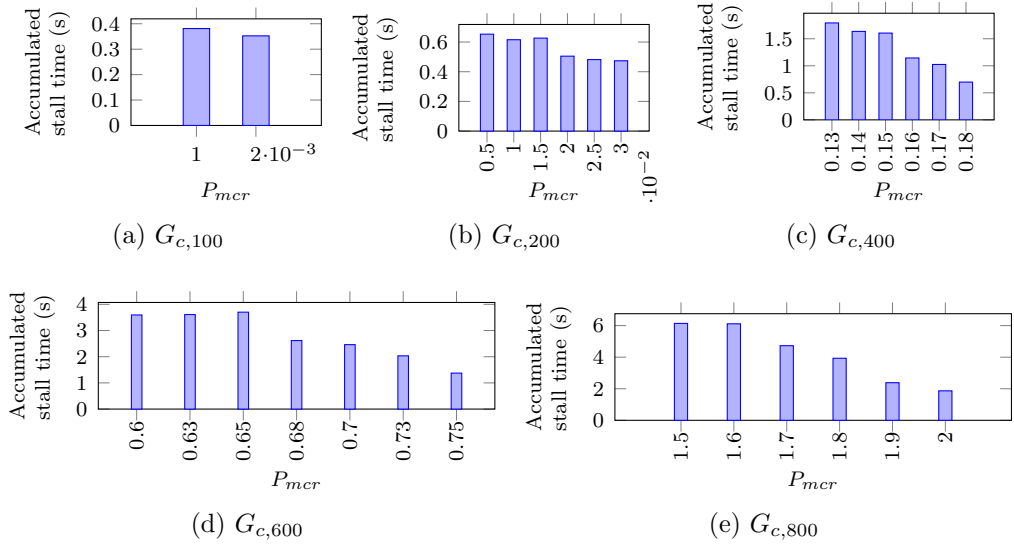

The total accumulated stall times for 100 MCR calls are shown in Fig.~\ref{fig:result3} for different sizes of $ G_c $. The results indicate that total stall time is inversely proportional to $ P_{mcr} $ as higher the $ P_{mcr} $ value, the number of MCRs served via the fallback method decreases. The decrease in the total stall times is small with smaller sizes of $ G_c $. This is expected since the oracle can perform DFS on $ G_c $ fast enough so that the stall time does not differ much from the ICD performed on $ G_{tc} $ and $ G^i_{tr} $. On the other hand, the difference with larger $ G_c $ is significant, for example, it is reduced by 70\% for $ G_{c,800} $ whereas only 0.075\% for $ G_{c,100} $, which are shown in Figs.~\ref{fig:r3-1} and \ref{fig:r3-5}, respectively. The decreases in the total stall times for other $ G_c $ are 37\% for $ G_{c,200} $, 61\% for $ G_{c,400} $ and also 61\% for $ G_{c,600} $ as shown in Figs.~\ref{fig:r3-2}, \ref{fig:r3-3} and \ref{fig:r3-4}.

\begin{figure}[tb]
\centering
\includegraphics[width=0.62\linewidth]{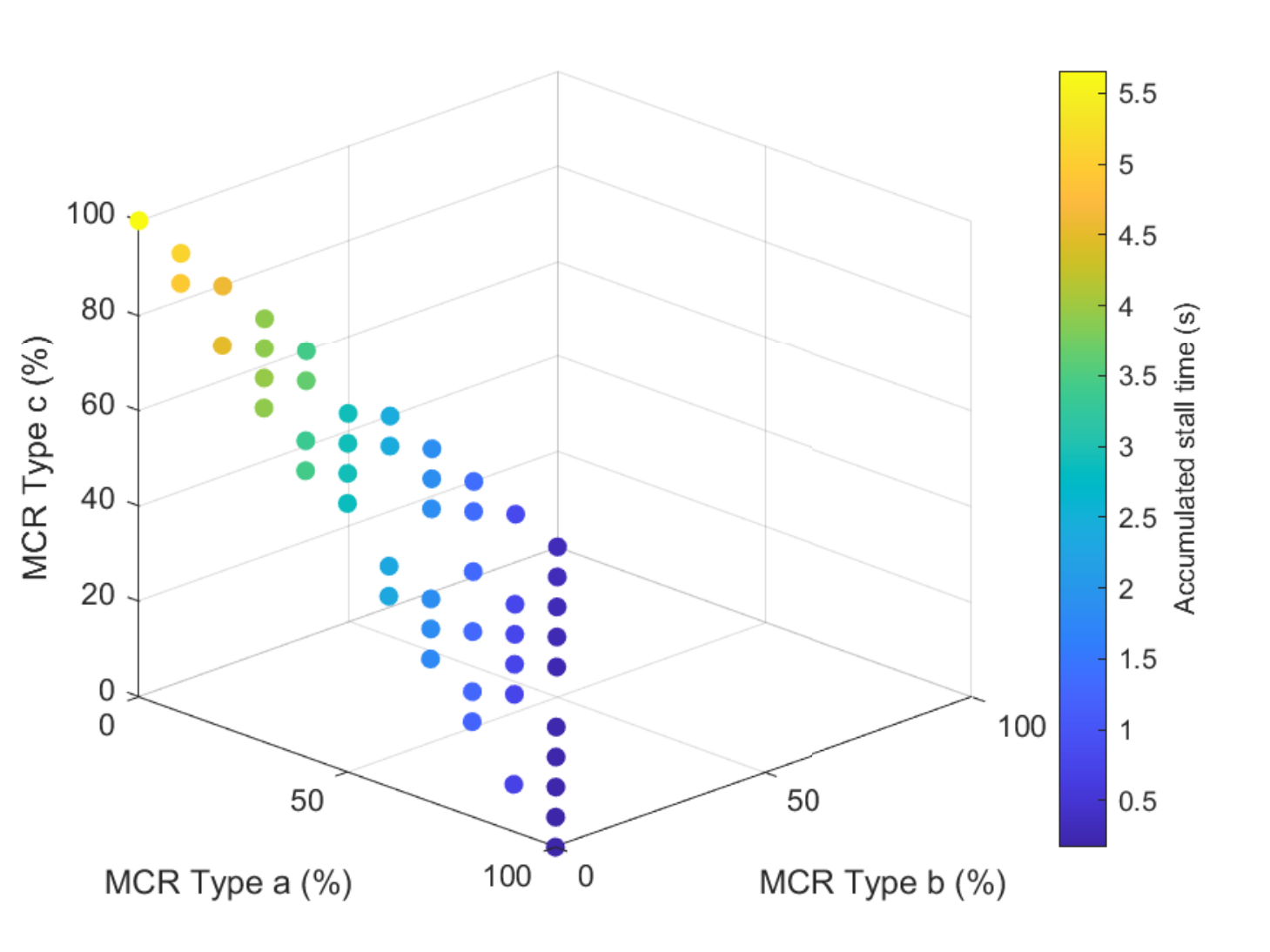}
\caption{Accumulated stall times for models with varying MCR types in 100 calls}
\label{fig:ratio}
\end{figure}

The previous experiment was carried out with 100 MCR calls with the types $ a $, $ b $ and $ c $ that are evenly distributed. The effect of different proportions of the MCR types on the stall time was investigated in the next experiment and the results are shown in Fig.~\ref{fig:ratio}. Each of the x, y and z axis represents the proportion of the MCR types that are sent to the oracle during the benchmark runs. The colours of the marks indicate the total accumulated stall times in seconds. This benchmark was run with $ G_c $ with the settings $ |V_c|=800 $, $ |E|=79900 $, $ |M_{mcr}| =10 $ and $ |E_{mcr}| =20 $. It is shown that the MCR types $ a $ and $ b $ did not significantly increase the total stall times whereas the time increased up to 5.5 seconds when all of the MCR calls were type $ c $.

\begin{figure*}[t!]
\centering
\subfloat[$ G_{c,400} $]{
\begin{tikzpicture}[baseline]
\begin{axis}[
height=3.0cm,
width=0.4\linewidth,
ylabel style={font=\scriptsize},
xlabel style={font=\scriptsize},
tick label style={font=\scriptsize},
ylabel=Percentage (\%),
xlabel=\# partitions,
xticklabels from table={results/Partition-400.csv}{part},
ybar=0.5pt,
bar width=4pt,
xtick=data,
ymin=0,
cycle list/Greys-3,
every axis plot post/.append style={fill, draw=black},
legend style={nodes={scale=0.6, transform shape},
             at={(0.5,1.1)},anchor=south},
legend columns=3,
legend image post style={scale=0.8},
enlarge x limits=0.25,
]
\addplot+ table[x=t,y=fall] {results/Partition-400.csv};
\addplot+ table[x=t,y=typec] {results/Partition-400.csv};
\addplot+ table[x=t,y=typeb] {results/Partition-400.csv};
\legend{Fallback,type-c,type-b}
\end{axis}
\end{tikzpicture}%
\hspace{\fill}
\begin{tikzpicture}[baseline]
\begin{axis}[
height=3.0cm,
width=0.4\linewidth,
tick label style={font=\scriptsize},
ylabel style={align=center,font=\scriptsize},
xlabel style={font=\scriptsize},
ylabel=Accumulated\\Stall time (s),
xlabel=\# partitions,
xticklabels from table={results/Partition-400.csv}{part},
ybar=0.5pt,
xtick=data,
bar width=8pt,
enlarge x limits=0.25,
ymin=0,
cycle list/Greys-3,
every axis plot post/.append style={fill, draw=black},
]
\addplot+ table[x=t,y=stall] {results/Partition-400.csv};
\end{axis}
\end{tikzpicture}
\label{fig:p1}
}

\subfloat[$ G_{c,600} $]{
\begin{tikzpicture}[baseline]
\begin{axis}[
height=3.0cm,
width=0.4\linewidth,
ylabel style={font=\scriptsize},
xlabel style={font=\scriptsize},
tick label style={font=\scriptsize},
ylabel=Percentage (\%),
xlabel=\# partitions,
xticklabels from table={results/Partition-600.csv}{part},
ybar=0.3pt,
bar width=3pt,
enlarge x limits=0.1,
xtick=data,
ymin=0,
cycle list/Greys-3,
every axis plot post/.append style={fill, draw=black},
legend style={nodes={scale=0.6, transform shape},
              at={(0.5,1.1)},anchor=south},
legend columns=3,
legend image post style={scale=0.8},
]
\addplot+ table[x=t,y=fall] {results/Partition-600.csv};
\addplot+ table[x=t,y=typec] {results/Partition-600.csv};
\addplot+ table[x=t,y=typeb] {results/Partition-600.csv};
\legend{Fallback,type-c,type-b}
\end{axis}
\end{tikzpicture}%
\hspace{\fill}
\begin{tikzpicture}[baseline]
\begin{axis}[
height=3.0cm,
width=0.4\linewidth,
tick label style={font=\scriptsize},
ylabel style={align=center,font=\scriptsize},
xlabel style={font=\scriptsize},
ylabel=Accumulated\\Stall time (s),
xlabel=\# partitions,
xticklabels from table={results/Partition-600.csv}{part},
ybar=0.5pt,
xtick=data,
bar width=8pt,
ymin=0,
cycle list/Greys-3,
every axis plot post/.append style={fill, draw=black},
]
\addplot+ table[x=t,y=stall] {results/Partition-600.csv};
\end{axis}
\end{tikzpicture}
\label{fig:p2}
}

\subfloat[$ G_{c,800} $]{
\begin{tikzpicture}[baseline]
\begin{axis}[
height=3.0cm,
width=0.4\linewidth,
ylabel style={font=\scriptsize},
xlabel style={font=\scriptsize},
tick label style={font=\scriptsize},
ylabel=Percentage (\%),
xlabel=\# partitions,
xticklabels from table={results/Partition-800.csv}{part},
ybar=0.3pt,
bar width=3pt,
enlarge x limits=0.12,
xtick=data,
ymin=0,
cycle list/Greys-3,
every axis plot post/.append style={fill, draw=black},
legend style={nodes={scale=0.6, transform shape},
              at={(0.5,1.1)},anchor=south},
legend columns=3,
legend image post style={scale=0.8},
xticklabel style={rotate=90},
]
\addplot+ table[x=t,y=fall] {results/Partition-800.csv};
\addplot+ table[x=t,y=typec] {results/Partition-800.csv};
\addplot+ table[x=t,y=typeb] {results/Partition-800.csv};
\legend{Fallback,type-c,type-b}
\end{axis}
\end{tikzpicture}%
\hspace{\fill}
\begin{tikzpicture}[baseline]
\begin{axis}[
height=3.0cm,
width=0.4\linewidth,
tick label style={font=\scriptsize},
ylabel style={align=center,font=\scriptsize},
xlabel style={font=\scriptsize},
ylabel=Accumulated\\Stall time (s),
xlabel=\# partitions,
xticklabels from table={results/Partition-800.csv}{part},
ybar=0.5pt,
xtick=data,
bar width=8pt,
ymin=0,
cycle list/Greys-3,
every axis plot post/.append style={fill, draw=black},
xticklabel style={rotate=90},
]
\addplot+ table[x=t,y=stall] {results/Partition-800.csv};
\end{axis}
\end{tikzpicture}
\label{fig:p3}
}
\caption{A set of experiments with different partition sizes for $ G_{tc} $}
\label{fig:part-r}
\end{figure*}
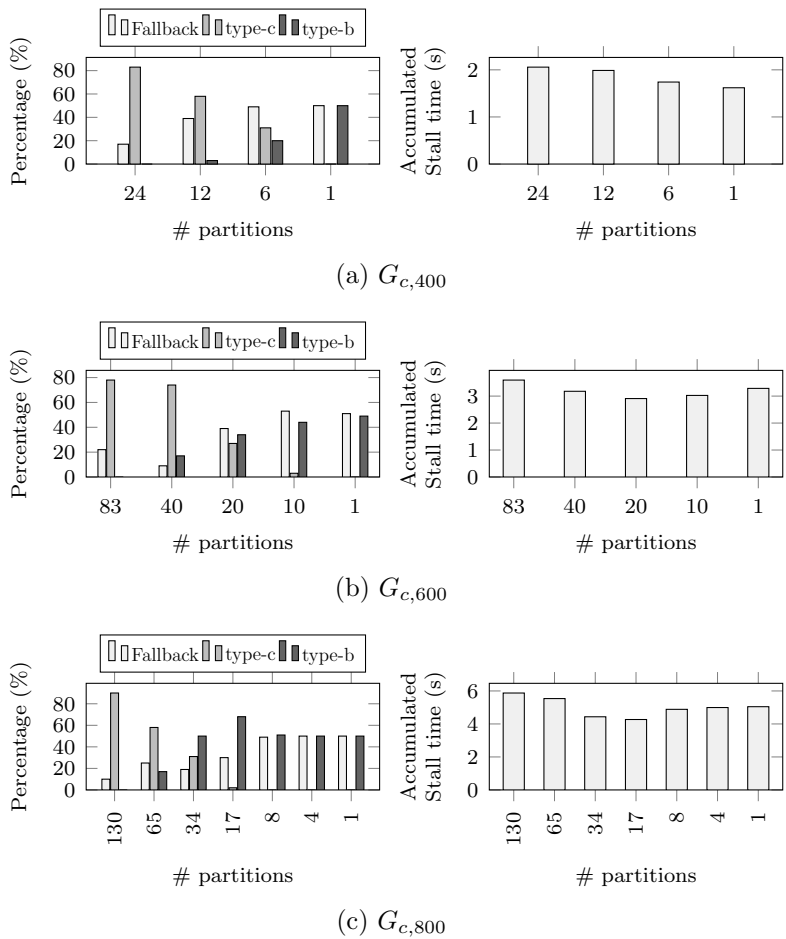

Fig.~\ref{fig:part-r} illustrates effects of various partition settings on $ G_{tc} $. The experiment was performed on the different sizes of partitions with a fixed MCR type $ c $ to observe how the partitions affect stall times on various $ G_c $ sizes. The first experiment on $ G_{c,400} $ in Fig.~\ref{fig:p1} shows conversions of MCR type $ c $ to type $ b $ as the number of partitions decreases. With a single large partition, about half of the MCRs were served by the oracle as type $ b $, see the left graph in Fig.~\ref{fig:p1}. On the other hand, the percentage of the DFS fallback method increased with the large partition because the time required to update $ G_{tr}^i $ and $ G_{tc} $ would also increase as explained in Section~\ref{sec:complexity}. We can observe a constant decrease in the accumulated stall time with increasing size (decreasing the number) of partitions with $ G_{c,400} $. The same trends can be observed for $ G_{c,600} $ and $ G_{c,800} $ as shown in Fig.~\ref{fig:p2} and Fig.~\ref{fig:p3}, respectively. However, it is shown that the accumulated stall times started to increase with large partitions for $ G_{c,600} $ and $ G_{c,800} $. This is due to the increased size of the $ G_{tr}^i $ that makes the time required for doing DFS on it becomes longer than those smaller $ G_{tr}^i $ in $ G_{c,400} $. To decrease stall time, this experiment suggests that one needs to partition $ G_{tc} $ in a way to convert MCRs of type $ c $ to type $ b $ as many as possible while avoiding to create too large partitions especially when the models within these partitions make frequent MCR calls. 


To apply our approach in a real-time setting, we foresee the use of a technique such as the probabilistic worst-case execution time (pWCET) analysis~\cite{cazorla2013proartis} to compute the worst-case stall time bound with certain degrees of confidence. In this technique, hardware or software components with hard-to-analyse timing behaviours are subject to statistical randomisation to reduce their computational complexity. Then, the rare event theories such as extreme value theory (EVT) can estimate the probability of maximum of the timing behaviours. In our case, such complexity arises from the distribution of MCR types on certain choices of partitions as well as their frequency that may or may not incur the DFS fallback method. Based on the observed behaviours of these parameters, we believe pWCET is applicable to the online ICD in the actual cyber-physical system.

\section{Experiment: The Workpiece Sorting System Case Study}\label{sec:wp-case}

This section presents the workpiece sorting system case study that is introduced earlier in Section~\ref{sec:mot}. The functional requirement of this experiment is to correctly place the workpieces into the bins in the presence of the stall time that is induced from the ICD. The system consists of \textit{twelve} pairs of controller and ejector plant models that sort incoming workpieces into one of the twelve bins. See Fig.~\ref{fig:ex}, which shows the first three ejectors placed along the conveyor belt. Additionally, there is also a controller that controls the speed of the conveyor belt. The operations performed by the machines upon the arrival of each workpiece consist of a sequence of mode changes. Our aim is to validate such mode changes via our online ICD approach without failing to sort incoming workpieces into correct bins due to the additional stall time that would create discrepancy between the model and the physical plant.

The oracle is added to include ICD capability for the mode changes in the controllers as well as the plant models for the ejectors. The ejector requires three mode changes, Idle-to-Push, Push-to-Pull and Pull-to-Idle, that push the workpiece into the bin. The corresponding MCRs are sent to the oracle in sequence when the workpiece is detected via the infra-red sensor at the ejector.

In addition to the system's operation described in Section~\ref{sec:mot}, we have extended the scenario with a constraint where the speed of the conveyor belt should decrease below the threshold of 0.5 unit/s before the ejector is able to push the workpiece into the bin. Furthermore, the real-time locations of the workpieces can only be detected via the infra-red sensors that are placed in line with the ejectors or at the entry point of the conveyor belt. 

The control of the conveyor belt's speed is done via the optimisation problem where the controller maximises the speed of the conveyor belt for a finite time-horizon. The stall time induced by the ICD lags the model's estimation on the physical location of the workpiece. Therefore the conveyor belt's speed might not decrease below the threshold before the physical workpiece reaches the ejector, if the stall time is too large. In this case, we consider the ejector incorrectly pushes the workpiece and hence fails. The lagged models can be re-synchronised with the physical system when the workpiece location is detected by one of the infra-red sensors. The models are developed using Simulink\footnote{https://www.mathworks.com/products/simulink.html}
where the controller for the conveyor belt is implemented using the MATLAB script whereas the SimEvent library is used for the conveyor's plant model. Models for the ejector device are implemented using the Stateflow diagram as well as the Simulink's continuous time blocks for capturing the ejector's physical movement. We have developed a Simulink-to-composite model converter from which $ G_c $, $ G_{tc} $ and $\{G^1_{tr},...,G^k_{tr}\} $ are generated.

\begin{figure}[t!]
\centering
\includegraphics[width=0.77\linewidth]{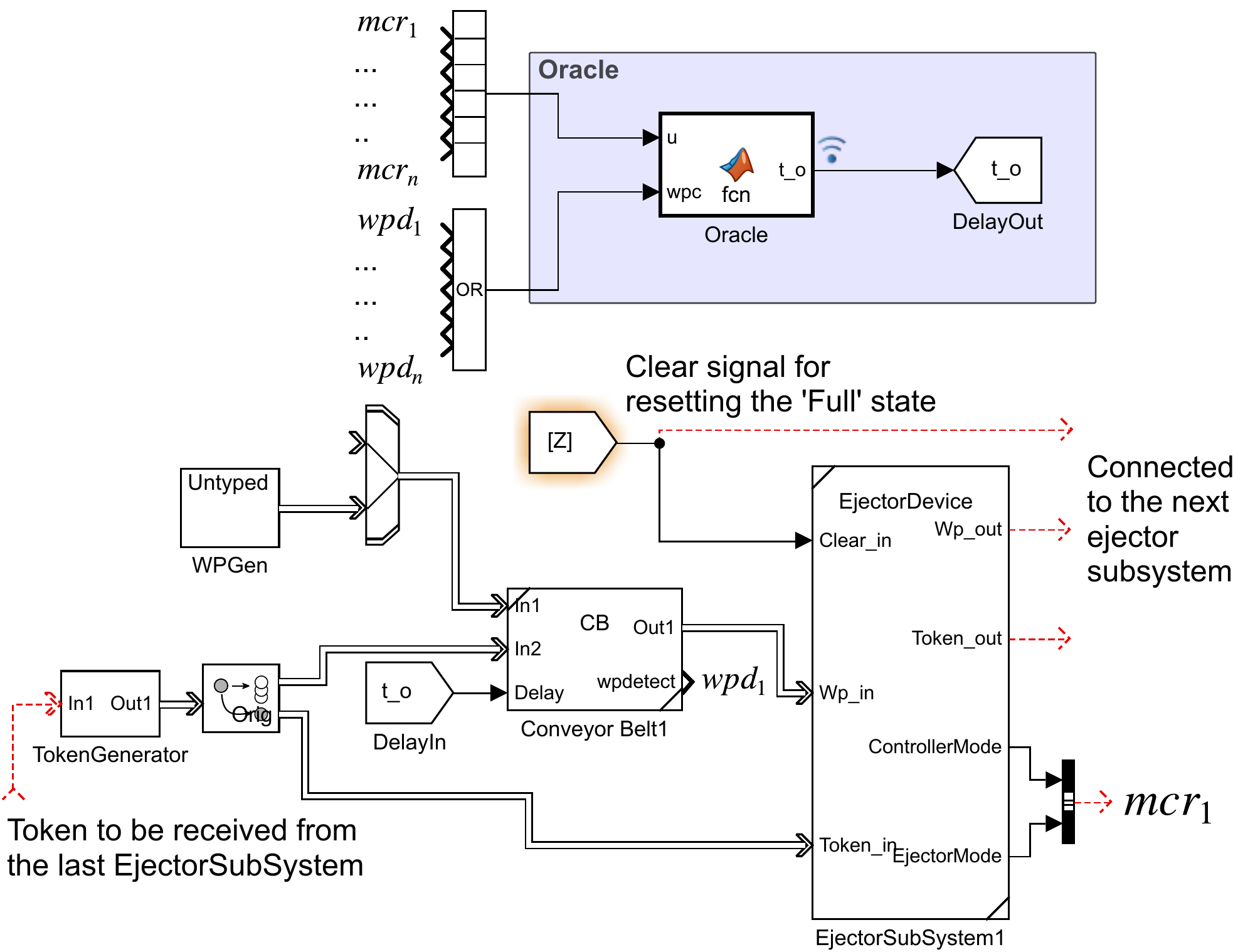}
\caption{Simulink top-level view of the workpiece sorting system}
\label{fig:sim-top}
\end{figure}

An overview of the workpiece sorting system is shown in Fig.~\ref{fig:sim-top}. The oracle annotated with a blue rectangle has concatenated input vectors of MCRs $ mcr_1,\dots,mcr_n $ where $ n $ is a total number of ejectors which is 12. An additional boolean typed input signal $ wpd $ indicating the exit of the workpiece from the conveyor belt is used for logging purpose. The output of the oracle is the analysis time for an MCR, which is fed to the input of the Conveyor Belt model for simulating the lagged delay. The block EjectorSubSystem1 has the controller and plant models for the ejector and the output red lines are connected to the adjacent ejector model, which is omitted due to space limitations. In this example, the current state of the controller and ejector models are used for constructing an MCR call. A token of the last ejector is returned to the EjectorSubSystem1 via TokenGenerator that implements token ring topology for deciding ejection order.

To generate $ G_c,G_{tc} $ and $ G_{tr} $ from the Simulink model shown in Fig.~\ref{fig:sim-top}, we traverse models based on their inter-connections using MATLAB's built-in Simulink programming model editing APIs and consider every Simulink model libraries as an individual model, i.e. we have $ \bigcup_{i=1}^k V^i_c = V_c $ where $ k $ is the total number of model libraries. In addition, we assume every model except the controller and ejector has fixed dependencies between every input and output ports. Each of the controller and ejector models have four and three modes, respectively, and we create dynamic input to output dependencies based on these modes as described in Section.~\ref{sec:mot}. The generated $ G_c $ has $ |V_c|=1409 $, $ |E_c| = 1441 $ and $ |M|=706 $.

\begin{figure}[!t]
\centering
\begin{tikzpicture}[baseline, every mark/.append style={scale=0.5}]
\pgfplotstableread[col sep=comma,]{results/run-dfs.csv}\tabledataa
\pgfplotstableread[col sep=comma,]{results/run-cau.csv}\tabledatab
\pgfplotstableread[col sep=comma,]{results/run-optimal.csv}\tabledatac
\pgfplotsset{
cycle list/Set1-4,
cycle multiindex* list={
mark list*\nextlist
Set1-4\nextlist
}}
\begin{axis}[
height=3.0cm,
width=0.8\linewidth,
xmin=0, xmax=37,
ylabel=Conveyor\\ speed (unit/s),
xlabel=Simulation time (s),
ylabel style={align=left,font=\scriptsize},
xlabel style={font=\scriptsize},
]
\addplot+[mark={none},Set1-A] table[x=time, y=v] {\tabledataa};

\addplot+[mark={none},Set1-B] table[x=time, y=v] {\tabledatab};

\addplot+[mark={none},Set1-C] table[x=time, y=v] {\tabledatac};

\addplot+[mark=x,semithick,ycomb,Set1-A] table[x=time1, y=o1] {\tabledataa};
\addplot+[mark=x,semithick,ycomb,Set1-A] table[x=time2, y=o2] {\tabledataa};
\addplot+[mark=x,semithick,ycomb,Set1-A] table[x=time3, y=o3] {\tabledataa};

\addplot+[mark=x,semithick,ycomb,Set1-B] table[x=time1, y=o1] {\tabledatab};
\addplot+[mark=x,semithick,ycomb,Set1-B] table[x=time2, y=o2] {\tabledatab};
\addplot+[mark=x,semithick,ycomb,Set1-B] table[x=time3, y=o3] {\tabledatab};

\addplot+[mark=x,semithick,ycomb,Set1-C] table[x=time1, y=o1] {\tabledatac};
\addplot+[mark=x,semithick,ycomb,Set1-C] table[x=time2, y=o2] {\tabledatac};
\addplot+[mark=x,semithick,ycomb,Set1-C] table[x=time3, y=o3] {\tabledatac};
\end{axis}
\end{tikzpicture}
\caption{Change in the conveyor belt speed when the instantaneous cycle detection performed only on $ G_c $ (red line), on $ G_{tc} $ (blue line) and $ G^i_{tr} $ and when the stall time is zero (green line)}
\label{fig:case}
\end{figure}
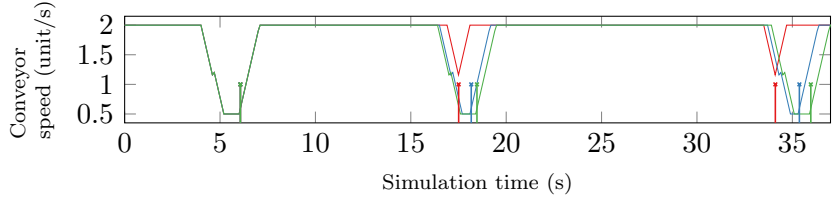

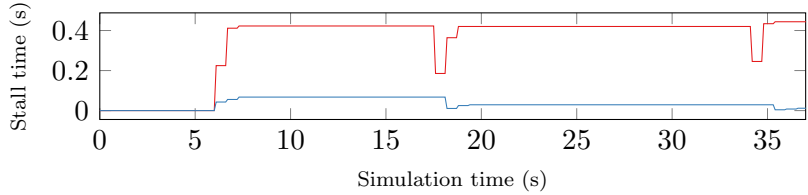
\begin{figure}[t!]
\centering
\begin{tikzpicture}[baseline, every mark/.append style={scale=0.5}]
\pgfplotstableread[col sep=comma,]{results/deltastall.csv}\tabledata
\pgfplotsset{
cycle list/Set1-4,
cycle multiindex* list={
mark list*\nextlist
Set1-4\nextlist
}}
\begin{axis}[
height=3.0cm,
width=0.8\linewidth,
xmin=0, xmax=37,
ylabel=Stall time (s),
xlabel=Simulation time (s),
ylabel style={align=left,font=\scriptsize},
xlabel style={font=\scriptsize},
]
\addplot+[mark={none},Set1-A] table[x=t1, y=st1] {\tabledata};
\addplot+[mark={none},Set1-B] table[x=t2, y=st2] {\tabledata};
\end{axis}
\end{tikzpicture}
\caption{Change in the accumulated stall times due to the instantaneous cycle detection}
\label{fig:accu}
\end{figure}

The results of the simulations are shown in Fig.~\ref{fig:case}. The plot consists of three simulation runs, indicated by the red, blue and green lines, until the completion of the first three ejector operations. The vertical lines are the instances when, in terms of simulation time, one of the infra-red sensors detected the workpiece. The red line that starts from 2 unit/s indicates the changes in the conveyor belt speed when the ICD was performed using DFS on $ G_c $ such as in~\cite{zhou2008causality}. The result shows the conveyor speed was correctly decreased below 0.5 unit/s for the first workpiece since no MCR was issued prior to this workpiece. However, the estimations of the next two workpieces arriving at the ejector were incorrectly computed due to the large stall time induced from the ICD on $ G_c $. As a result, the model simulation significantly lagged behind the physical system's operation and this is indicated by the vertical red line at around 34.11 s that appeared when the conveyor belt speed was above 0.5 unit/s threshold. On the other hand, the blue line shows the result of our approach where the conveyor belt speed was correctly reduced below the threshold before the workpiece reached the ejector. We have partitioned the models in a way that the oracle could utilise $ G_{tc} $ or $ G^i_{tr} $ for any MCRs. The ideal case is shown as the green line, when the stall time was zero, for comparison purposes. The workpiece detection times (vertical lines) are different depending on the analysis methods as shown from the second ejector operation. This is because the conveyor belt speeds were reduced at different simulation times due to the varying stall times depending on the ICD methods.

The changes in the accumulated stall times for the results presented in Fig.~\ref{fig:case} are shown in Fig.~\ref{fig:accu}. It is clearly shown that the stall time increased significantly up to 0.434 s when the ICD was done only on $ G_c $ whereas it is only up to 0.067 s when our oracle-based approach was used. The models are re-synchronised with the physical system when the workpiece is detected by the sensor. This is indicated by the stall time that drops at around 17.5 s and 57 s. At the same time, the ejector needs to perform the push operation that again induced additional stall time, and therefore the adjusted stall time did not drop to zero as shown in the figure.

\section{Further Discussion on the Use-Case Scenario}\label{sec:dis}

With the development of Functional Mock-up Interface (FMI) standard~\cite{broman2013determinate}, many modelling tools start to support exporting and importing third-party models encapsulated in a standard container called the Functional Mock-up Unit (FMU). FMI standard defines how IO dependencies can be embedded in FMU. Nevertheless, the standard only supports fixed dependencies for all possible modes in FMU. For example, \cite{website:model-aircraft} demonstrates the use of FMU that encapsulates thermal models of the more-electric aircraft. This model, when flattened and consisting of only basic atomic blocks, has 1668 ports ($ |V_c| $) and 1492 signal connections ($ |E_c| $). The model has a delay block in front of the FMU in order to break potential instantaneous cycles. Such delay blocks might not be needed or even result in undesired behaviour when the models are executed in the SR semantics and need to be synchronised in the exact SR tick (Section~\ref{ssec:sr}). Our technique can improve the dependency analysis for the models that employ FMU especially with the upcoming FMI version (3.0) that supports co-simulation of models with discrete events. 

Several hybrid system models are presented in~\cite{carloni2006languages} with discussions on the creation of instantaneous cycles. For example, three-point masses moving on a flat surface that collides with each other (discrete reset) during which instantaneous cycles can occur in the model. Another example called the full wave rectifier is presented in~\cite{carloni2006languages} where an instantaneous cycle is created when the load to the system is replaced with a pure resistive load. Again, the existence of mode switches in this model makes the static analysis on the component-based models difficult. Alternatively, an online technique can alleviate this problem and our method targets reducing potential discrepancies between the model and physical system due to the ICD. The three-point mass example consists of approximately 70 ports and 35 signal connections while 21 ports and 20 signal connections for the rectifier example.

In \cite{hofbaur2006causal}, an example of fault tolerant micro-chemical plant was introduced where a chemical substance produced from multiple reactants has to be mixed before it reaches the final output tank. The fault scenarios that impair multiple pumps and heat exchangers require reconfiguration of mixers so the substance can be correctly rerouted to the output tank. In this case, mixers change their input to output port dependencies that can result in instantaneous cycle. Therefore, our online approach is also applicable in this scenario when the number of combined modes for the system is significantly large ($ 10^6 $) that static analysis infeasible as indicated by the authors. Furthermore, when the models are used online to quickly adapt to faults and provide possible reroutes, reducing stall time from our ICD technique would be beneficial. While the authors presented a simplified version of the plant that has 2 reactants, 20 valves, 4 pumps and 2 heat exchangers, which result in approximately 54 ports and 43 IO connections, the size of real system is typically much larger.

\section{Conclusions and Future Work}\label{sec:concl}
This paper addressed the problem of online ICD for the models with mode-dependent input and output dependencies that change during runtime. The method utilises an oracle that determines whether the models can change their internal modes at each synchronous tick boundary. The method can be applied to various modelling formalisms such as the synchronous reactive model of computation if there is a notion of synchronisation point for communication with the oracle.
The oracle maintains three types of data-structures: a composite graph and its transitive closure as well as transitive reduction, which are adaptively chosen to reduce the stall time of the model simulation induced by the online ICD. The experimental results showed the oracle can provide an answer to the mode update queries within an acceptable stall time frame. We also demonstrated the industrial use-case of our technique via the workpiece sorting system.
For future work, we would like to further investigate the impacts of various data-structures for the dynamic transitive closure with varying query and update complexities. 

\section*{Acknowledgement}

This work was supported by Delta-NTU Corporate Lab for Cyber-Physical Systems with funding support from Delta Electronics Inc. and the National Research Foundation (NRF) Singapore under the Corp Lab@University Scheme.




\bibliographystyle{elsarticle-num} 
\bibliography{bibliography}
\end{document}